\documentclass[aps,prb,twocolumn,longbibliography,floatfix,preprintnumbers,superscriptaddress,amsmath,amssymb]{revtex4-2}

\usepackage{xcolor,bm,lineno,multirow,times}
\usepackage{physics}
\usepackage{amsmath}
\usepackage{rotating}
\usepackage{verbatim}
\usepackage{graphicx,tabularx}
\usepackage[multiple]{footmisc}
\usepackage[sort&compress]{natbib}
\usepackage[T1]{fontenc}
\usepackage{soul}
\usepackage{dcolumn}
\usepackage{bm}
\usepackage[version=4]{mhchem}
\usepackage{booktabs}
\usepackage{gensymb}
\usepackage{color}
\usepackage[colorlinks=true,urlcolor=blue,linkcolor=blue,citecolor=blue]{hyperref}
\usepackage[title]{appendix}
\usepackage{orcidlink}
\usepackage{todonotes}
\usepackage{glossaries}
\newacronym{ED}{ED}{Exact diagonalization}
\newacronym{DMRG}{DMRG}{density matrix renormalization group} 
\newacronym{QMC}{QMC}{Quantum Monte Carlo} 
\newacronym{LSWT}{LSWT}{linear spin wave theory} 
\newacronym{GLSWT}{GLSWT}{generalized linear spin wave theory} 
\newacronym{LL}{LL}{Landau-Lifshitz} 
\newacronym{GLL}{GLL}{generalized Landau-Lifshitz} 
\newacronym{DSSF}{DSSF}{dynamical spin structure factor} 

\usepackage{floatrow}

\newcommand{\mb}[1]{\mathbf{#1}}

\begin{document}
\title{Classical dynamics of the antiferromagnetic Heisenberg $S=1/2$ spin ladder}
\author{David A. Dahlbom\orcidlink{0000-0002-0221-5086}}
\email{dahlbomda@ornl.gov}
\affiliation{Department of Physics and Astronomy, University of Tennessee, Knoxville, TN 37996, USA}
\affiliation{Neutron Scattering Division,  Oak Ridge National Laboratory, Oak Ridge, TN 37830, USA}
\author{Jinu Thomas\orcidlink{0000-0003-4818-6660}}
\affiliation{Department of Physics and Astronomy, University of Tennessee, Knoxville, TN 37996, USA}
\affiliation{Institute for Advanced Materials and Manufacturing, The University of Tennessee, Knoxville, TN 37996, USA} 
\author{Steven Johnston\orcidlink{0000-0002-2343-0113}}
\affiliation{Department of Physics and Astronomy, University of Tennessee, Knoxville, TN 37996, USA}
\affiliation{Institute for Advanced Materials and Manufacturing, The University of Tennessee, Knoxville, TN 37996, USA} 
\author{Kipton Barros\orcidlink{0000-0002-1333-5972}}
\affiliation{Theoretical Division and CNLS, Los Alamos National Laboratory, Los Alamos, NM 87545, USA}
\author{Cristian D. Batista\orcidlink{0000-0003-1667-3667
}}
\affiliation{Department of Physics and Astronomy, University of Tennessee, Knoxville, TN 37996, USA}

\begin{abstract}
We employ a classical limit grounded in SU(4) coherent states to investigate the temperature-dependent dynamical spin structure factor of the $S=1/2$ ladder consisting of weakly coupled dimers. By comparing the outcomes of this classical approximation with density matrix renormalization group and exact diagonalization calculations in finite size ladders, we demonstrate that the classical dynamics offers an accurate approximation across the entire temperature range when the interdimer coupling is weak and a good approximation in the high temperature regime even when the interdimer coupling is strong. This agreement is achieved after appropriately rescaling the temperature axis and renormalizing expectation values to satisfy a quantum sum rule, following D. Dahlbom \textit{et al}. [Phys. Rev. B {\bf 109}, 014427 (2024)]. We anticipate the method will be particularly effective when applied to 2D and 3D lattices composed of weakly-coupled dimers, 
situations that remain challenging for alternative numerical methods.
\end{abstract}

\maketitle

\section{Introduction}

Calculating the dynamical correlation functions of interacting quantum spin systems at arbitrary temperatures represents a pivotal yet unresolved challenge in quantum many-body theory.  Such calculations are essential for extracting the underlying model Hamiltonian of a material because dynamical susceptibilities, linked to two-point correlation functions,  can be measured via a range of spectroscopic and resonance experiments. Moreover, dynamical susceptibilities serve as a window into the nature of collective modes that are the ``fingerprints'' of each state of matter. 

In general, the computational cost of calculating dynamical spin-spin correlation functions in thermal equilibrium grows exponentially in the  number of spins. This exponential complexity has spurred the development of various approximation schemes and numerically exact approaches. However, state of the art numerical techniques for simulating quantum many-body problems still have severe limitations. \gls*{ED} techniques~\cite{Gagliano88} are confined to small clusters. The \gls*{DMRG} \cite{White92,White1993density,Hallberg06} is applicable to one-dimensional systems or narrow ribbons, but it has severe limitations for studying two- and three-dimensional magnets. Tensor networks are capable of handling certain 2D systems, but efforts to compute dynamical spin structure factors with this technique are recent and based on single-mode approximations at $T=0$~\cite{Verresen19,Chi22,Lin22,Sherman23,Xu24,Gao24}. \gls*{QMC} techniques are most effective at low-temperatures for the subset of models without a sign-problem and, even for this subset, calculating dynamical correlation functions from noisy data is challenging due to the fundamentally ill-posed task of analytical continuation from the Matsubara domain to real frequencies~\cite{Hirsch83, JARRELL96, Sandvik98, Sandvik16,Goulko17,Nichols22,Li23,Nogaki23,SHAO2023}.
Given these limitations, efficient numerical  techniques capable of providing a good approximation to exact temperature-dependent dynamical correlation functions are of general interest to the quantum many-body community.

In this study, we employ a generalization of the classical limit applied to quantum spin systems~\cite{Zhang21, Dahlbom22, Dahlbom22b, Dahlbom23} to investigate the approximate spin dynamics of an antiferromagnetic $S=1/2$ quantum ladder featuring a non-magnetic ground state. 
Conventional classical approximations obtained by taking the large-$S$ limit encounter challenges when dealing with coupled dimer systems exhibiting predominant intra-dimer antiferromagnetic interactions. This limitation arises from the inability to represent the singlet ground state of an isolated dimer (non-magnetic solution) as a product of two SU(2) coherent states (magnetic solution). More broadly, non-magnetic ground states of quantum paramagnets that are composed of interconnected entangled units similarly defy approximation based on direct products of SU(2) coherent states, which assign a classical dipole of fixed magnitude ($S$) to each spin within the unit. As our discussion will reveal, a classical limit based on SU($N$) coherent states, with $N$ denoting the number of levels within each entangled unit, offers a suitable approximation for these systems by effectively capturing their intra-unit entanglement. 

The classical theory based on SU($N$) states corresponds in a precise sense to certain ``multiflavor'' spin wave methods in the same way that the Landau-Lifshitz equations correspond to traditional linear spin wave theory. To make this connection clear, we review the well-known multiflavor bosonization procedure for the spin ladder in the language of SU($N$) coherent states, emphasizing how the control parameter used in the development of the spin wave expansion is precisely the same one that is sent to infinity when deriving the generalized classical limit. Just as the Landau-Lifshitz dynamics captures important nonlinearities and facilitates finite-temperature simulation of spin systems that are well-described with a dipolar approximation, the classical limit described here offers the same benefits for systems with non-magnetic ground states.

The insights gained from this study extend beyond quantum ladders and can be applied to larger ``entangled units,'' such as trimers or tetrahedra, whenever these units contain spins with prevailing antiferromagnetic interactions. The rationale for selecting a one-dimensional (1D) ladder for this investigation is multifaceted. First, 1D systems provide a stringent test for classical approximations. Their low coordination number amplifies the extent of the ground state entanglement. The $S=1/2$ dimers constituting the entangled units of the ladder exhibit a robust singlet character when the intra-rung exchange parameter ($J$) dominates over the inter-rung exchange parameter ($J'$) (see Fig.~\ref{fig:ladder}). However, the inter-dimer entanglement experiences rapid growth with the ratio $J'/J$ because of the increased weight of singlet configurations along the legs. These length-wise dimers resonate with each other and render classical approximations increasingly inaccurate. Secondly, the advantageous low-coordination number of 1D systems enables us to accurately compute the exact spin dynamics at arbitrary temperatures for ladder sizes that surpass their entanglement length scale and encompass a substantial number of distinct momenta along the leg direction, facilitating the extrapolation of dynamics to the thermodynamic limit. 

It is generally accepted that most quantum mechanical systems exhibit a quantum-to-classical crossover upon increasing temperature. It is not always clear, however, what classical limit to take. An important outcome of this work is that, above a certain crossover temperature  $T^*$, the generalized classical dynamics offer a remarkably accurate approximation to the \gls*{DSSF}. Leveraging this classical approximation, we can effectively determine the Hamiltonian parameters by fitting the exact \gls*{DSSF} with the approximated outcome in the infinite temperature regime. This finding holds significant practical implications, particularly in extracting models from high-temperature inelastic neutron scattering data.

Beyond parameter estimation, expanding the set of classical theories available to describe spin systems serves a more fundamental purpose in evaluating the essential ``quantumness'' of an observed phenomenon \cite{Zhang_2019, Bai19, samarakoon2017comprehensive}. A broad scattering response, for example, could be a consequence of weakly-confined fractionalized excitations or nonlinear magnon-magnon interactions. Since the classical limit in some sense captures magnon-magnon interactions out to arbitrary order, examining the finite-temperature dynamics of a classical model offers a straightforward way to assess the contribution of nonlinearities \cite{winter2017breakdown, Franke2022}. As we observe in this study, the failure of the traditional classical limit to capture the essential nonlinearities of a spin system does not preclude the possibility that a more appropriate classical limit would succeed. 

This work is organized as follows. Sec.~\ref{sec:classical_theory} introduces the $S=1/2$ Heisenberg ladder and a classical limit based on the representation of dimers as SU(4) coherent states. Sec.~\ref{sec:S(q,w)} explains how this formalism can be used to calculate the \gls*{DSSF} classically and introduces the classical-to-quantum correspondence factor for obtaining an approximation to the quantum mechanical \gls*{DSSF}. In Sec.~\ref{sec:T=0_DMRG_comparisons}, a generalized spin wave calculation corresponding to the classical theory is developed, expanding around a product of SU(4) coherent states. The results are compared against those obtained with the \gls*{DMRG} method. In Sec.~\ref{sec:finite_T}, we return to the purely classical dynamics to calculate the finite temperature \gls*{DSSF} and compare against the results of \gls*{ED}. Our conclusions are presented in Sec.~\ref{sec:concl}.


\section{The classical limit of the $S=1/2$ spin ladder in coherent states of SU(4)\label{sec:classical_theory}}
\begin{figure}[t]
   \centering
    \includegraphics[width=1.0\columnwidth]{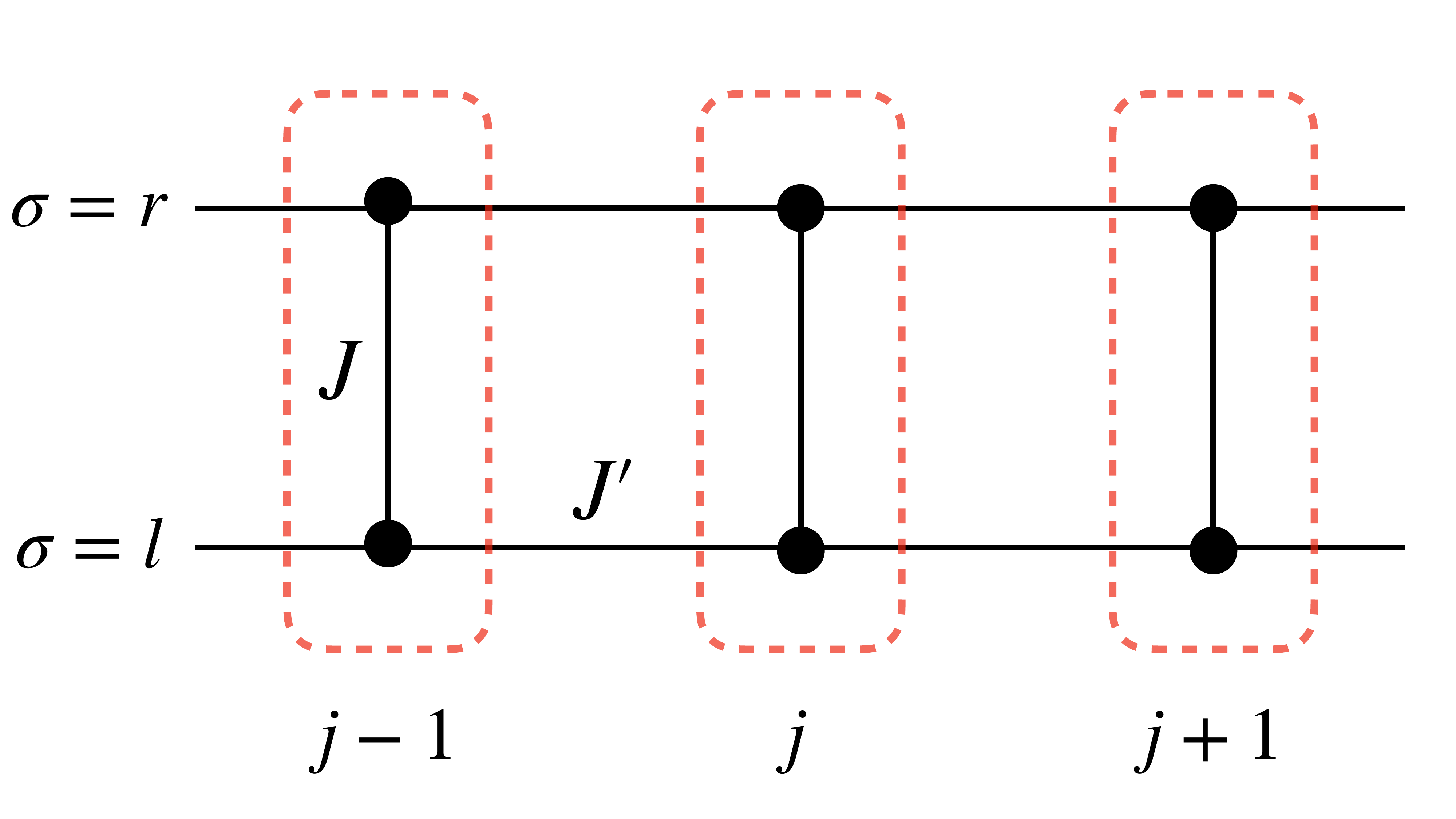}
    \caption{Illustration of the $S=1/2$ spin ladder. We assume $J^\prime/J < 0.5$, leading to a singlet ground state on each rung. While a singlet state cannot be represented as the product of two SU(2) coherent states, it may be represented as a single SU(4) coherent state. We develop a classical theory based on assigning one such coherent state to each rung (dashed red boxes), each containing two $S=1/2$ spins (black circles).
    }
\label{fig:ladder}
\end{figure}

We will consider the $S=1/2$ Heisenberg spin ladder Hamiltonian
\begin{equation}
    \mathcal{\hat{H}} = J \sum_{j}  \hat{S}_{j, l}^{\beta} \hat{S}_{j, r}^{\beta} + J^{\prime} \sum_{j} \left( \hat{S}_{j, l}^{\beta} \hat{S}_{j+1, l}^{\beta} + \hat{S}_{j, r}^{\beta} \hat{S}_{j+1, r}^{\beta} \right),
    \label{eq:hamiltonian}
\end{equation}
where $j$ is the rung index, $l$ indicates the left  leg of the ladder, and $r$ the right, as illustrated in Fig.~\ref{fig:ladder}. Summation over repeated Greek indices is assumed. Throughout this study, we will take $J$ as our unit of energy.

A classical limit of a quantum system can be viewed as a systematic approach to neglecting entanglement. In the context of a spin system, this traditionally takes the form of 
neglecting entanglement between individual spins. Formally, this is achieved by restricting consideration to product states
\begin{equation}
\left\vert \Omega \right\rangle = \bigotimes_{j,\sigma} \left\vert \Omega_{j,\sigma} \right\rangle,
\label{eq:product_state_su2}
\end{equation}
where $\sigma=l,r$ and each factor $|\Omega_{j,\sigma}\rangle$ is an SU(2) coherent state, which will be described below.
 Using this decomposition, one derives an approximate Hamiltonian that evolves products of SU(2) coherent states into other products of SU(2) coherent states, i.e., a Hamiltonian that generates group actions formed by
 the tensor product of SU(2) group transformations acting on each site $\left(j,\sigma\right)$. The corresponding
 Lie algebra, in terms of which the approximate Hamiltonian must be expressed, is the direct sum of local 
 $\mathfrak{su}\left(2\right)$ Lie algebras. The procedure for deriving this approximate Hamiltonian is described below.

For intuition's sake, we note that an SU(2) coherent state may always be identified with a pure state of a two-level system. Furthermore, the space of possible observables on a 2-level system  is
spanned by the identity and the  dipole operators $\hat{S}_j^x, \hat{S}_j^y$, and $\hat{S}_j^z$ that generate the local $\frak{su}(2)$ Lie algebra:
\begin{equation}
[\hat{S}_j^{\mu}, \hat{S}_j^{\nu}]   = \mathrm{i} \epsilon_{\mu \nu \eta} \hat{S}_j^{\eta}.
\end{equation}
A consequence of this is that any SU(2) coherent state may be put into one-to-one correspondence
with the expectation values of these observables through the mapping
\begin{equation}
    s^{\alpha}_{j,\sigma} = \left\langle\Omega_{j,\sigma}\left\vert\hat{S}_{j,\sigma}^\alpha\right\vert\Omega_{j,\sigma}\right\rangle
\end{equation}\footnote{The inverse mapping,
from a set of expectation values, $s^\alpha$, to an SU(2) coherent state, is constructed by finding the highest-weight eigenvector of $s^\alpha\hat{S}^\alpha$.}. This correspondence is simply the familiar Bloch sphere construction and is the underlying reason that the traditional classical limit yields a theory of dipoles. We emphasize that an approach based on SU(2) coherent states is able to represent classically the state of a 2-level system on each site, such as an $S=1/2$ spin, and it is appropriate whenever entanglement between different 2-level sites may be disregarded.

The traditional classical Hamiltonian is derived 
by replacing each operator with its expectation value in such a product state.
The expectation is evaluated in the large-$S$ limit [where $S$ labels irreps of SU(2)]. In this limit,
spin operators may be replaced with their expectation values, yielding
\begin{equation}
\begin{alignedat}{2}
H_{\mathrm{SU(2)}} &= \lim_{S\to\infty} \left\langle \Omega \left\vert \hat{\mathcal{H}} \right\vert \Omega \right\rangle \\
    &= J \sum_{j}  s_{j, l}^{\beta} s_{j, r}^{\beta} + J^{\prime} \sum_{j} \left( s_{j, l}^{\beta} s_{j+1, l}^{\beta} + s_{j, r}^{\beta} s_{j+1, r}^{\beta} \right). \\
\end{alignedat}
\label{eq:hamiltonian}
\end{equation}
The dynamics may be derived by considering the Heisenberg equations of motion
\begin{equation}
\mathrm{i}\hbar\frac{d\hat{S}_{j,\sigma}^{\alpha}}{dt} =\left[\hat{S}_{j,\sigma}^{\alpha},\hat{\cal{H}}\right]
=\mathrm{i}\epsilon_{\alpha\beta\gamma}\frac{\partial\mathcal{\hat{H}}}{\partial\hat{S}_{j,\sigma}^{\beta}}\hat{S}_{j,\sigma}^{\gamma}.
\end{equation}
We use the operator derivative as a shorthand for an expression that becomes exact in the large-$S$ limit. See \cite{Zhang21, Dahlbom22} and the appendix of \cite{Dahlbom23} for details. 
Evaluating the expectation value in coherent states in the large-$S$ limit, we find
\begin{equation}
\frac{ds_{j,\sigma}^{\alpha}}{dt}= \frac{\epsilon_{\alpha\beta\gamma}}{\hbar} \frac{\partial H_{\rm SU(2)}}{\partial s_{j,\sigma}^{\beta}} s_{j,\sigma}^{\gamma}.
\label{eq:landau-lifshitz}
\end{equation}
The result is the familiar \gls*{LL} equation~\cite{Lakshmanan11}.

An alternative classical limit may be taken by retaining the entanglement on the two sites bonded on a rung while still neglecting entanglement between rungs. Instead of restricting to products of SU(2) coherent states, as in Eq.~\eqref{eq:product_state_su2},
we will restrict to products of SU(4) coherent states
\begin{equation}
    \left\vert \Psi \right\rangle = \bigotimes_j \left\vert \Psi_j \right\rangle,
\end{equation}
where $j$ indexes rungs. Just as an SU(2) coherent state may be identified with a pure state in a 2-dimensional Hilbert space, an SU(4) coherent state may be identified with a pure state in a 4-dimensional Hilbert space like those that exist on a bond joining two $S=1/2$ spins. Importantly, the set of all such pure states is capable of representing
any entangled state between two $S=1/2$ spins. Furthermore, just as an SU(2) coherent state
may be identified with the expectation values of the spin operators [the Lie algebra $\mathfrak{su}(2)$], an SU(4) coherent state
may be put into one-to-one correspondence with the expectations of a complete set of observables acting on a 4-dimensional Hilbert space
[the Lie algebra $\mathfrak{su}(4)$]. We will call these observables $\hat{T}^\alpha$, where $\alpha$ runs from 1 to 15 [the dimension of $\mathfrak{su}(4)$].

A convenient choice of generators is given by
\begin{equation}
\begin{alignedat}{4}
\hat{T}^{1} &= \hat{S}^x_{l} & \quad \hat{T}^{2} &= \hat{S}^y_{l} & \quad \hat{T}^{3} &= \hat{S}^z_{l} \\ 
\hat{T}^{4} &= \hat{S}^x_{r} & \quad \hat{T}^{5} &= \hat{S}^y_{r} & \quad \hat{T}^{6} &= \hat{S}^z_{r} \\ 
\hat{T}^{7} & = 2\hat{S}^x_{l}\hat{S}^x_{r} & \quad \hat{T}^{8} &=  2\hat{S}^y_{l}\hat{S}^y_{r}  & \quad \hat{T}^{9} &=  2\hat{S}^z_{l}\hat{S}^z_{r}\\
\hat{T}^{10} &= 2\hat{S}^y_{l}\hat{S}^z_{r} & \quad \hat{T}^{11} &= 2\hat{S}^z_{l}\hat{S}^y_{r} & \quad \hat{T}^{12} &= 2\hat{S}^z_{l}\hat{S}^x_{r}\\
\hat{T}^{13} &= 2\hat{S}^x_{l}\hat{S}^z_{r} & \quad  \hat{T}^{14} &= 2\hat{S}^x_{l}\hat{S}^y_{r} & \quad \hat{T}^{15} &= 2\hat{S}^y_{l}\hat{S}^x_{r}, 
\end{alignedat}
\label{eq:generators}
\end{equation}
where each $\hat{S}_{\sigma}^\alpha$ is given in the $S=1/2$ representation. These generators are orthonormal in the 
sense that $\mathrm{Tr}\,\hat{T}^\alpha\hat{T}^\beta = \delta_{\alpha\beta}$ and complete in
that any observable on a 4-dimensional Hilbert space may be written as $c_0 \hat{1}+ c_\alpha \hat{T}^\alpha$ for real coefficients $c_0$ and $c_\alpha$.
 We emphasize that the specific choice of generators is somewhat arbitrary; they need only constitute a faithful representation of the Lie algebra $\mathfrak{su}(4)$. In particular, the familiar decomposition of the observables of a single SU(4) spin into irreps of SO(3) (multipolar moments) does not apply, as we are here describing two $S=1/2$ spins rather than a single $S=(N-1)/2=3/2$ spin. The generators of Eq.~(9) are well-adapted to the form of the ladder Hamiltonian, making later calculations simpler. They also yield a particularly sparse set of structure constants, and, consequently, relatively simple classical equations of motion.
 Relevant observables may be constructed by symmetrizing or anti-symmetrizing the bilinear forms ($T^{\alpha}/2$ with $\alpha=7,\dots,15$).
The  symmetric form  ${\bm S}_l \cdot {\bm S}_r$, which is a scalar under global spin rotations, measures the singlet (eigenvalue -3/4) versus triplet character (eigenvalue 1/4) of the dimer. The symmetric form $Q^{\alpha \beta}= \hat{S}^\alpha_{l}\hat{S}^\beta_{r} + \hat{S}^\beta_{l}\hat{S}^\alpha_{r} - \frac{2}{3} \hat{\bm S}_l \cdot \hat{\bm S}_r$ corresponds to a bond quadrupolar moment, and the anti-symmetric form $\hat{\bm S}_l \times \hat{\bm S}_r$ is proportional to the spin current density on the bond. 

With this choice of generators, we introduce the classical ``color field''  
\begin{equation}
n^\alpha_j = \left\langle \Psi \left\vert \hat{T}^\alpha \right\vert \Psi \right\rangle,
\label{eq:color_field}
\end{equation}
which is a 15-element real vector. Its first six elements correspond to the expected dipoles on each of the two sites comprising rung $j$. Its remaining elements can be interpreted physically in terms of the bilinear forms just described. 
Formally, the expectation values  
are evaluated in the $M \to \infty$ limit, where the integer $M$ labels completely symmetric irreducible representations (irreps) of SU(4)  \footnote{A representation of SU($N$) is uniquely specified by the $N-1$ eigenvalues of the Cartan subalgebra, $\{\lambda_1,\ldots,\lambda_{N-1}\}$. We are specifically referring to the completely symmetric irreps with a single non-zero eigenvalue: $\lambda_1=M$ and $\lambda_n=0$ for $n>1$. Note that \cite{muniz2014generalized} uses $M$ to label these irreps, while other recent publications \cite{Zhang21, Dahlbom22b, Dahlbom23} use the notation $\lambda_1$ to indicate the same thing.}.  Since operators commute to leading order in $M$, this involves replacing operators with their expectation values. In particular, this leads to a factorization rule, $\langle \hat{T}^{\alpha} \hat{T}^{\beta} \rangle = \langle \hat{T}^{\alpha} \rangle \langle \hat{T}^{\beta} \rangle $. Application of this factorization rule to the quadratic Casimir satisfied by the generators [Eq.~\eqref{eq:generators}] leads to the conclusion that the norm of the color field is constant, just as a dipole has a fixed magnitude $S$ in the classical limit based on SU($2$) coherent states. This will be discussed further in Sec.~\ref{sec:finite_T}.

Since our Hamiltonian Eq.~\eqref{eq:hamiltonian} is linear in terms of the generators of Eq.~\eqref{eq:generators} on each site, it is not necessary to apply the factorization rule to derive the classical Hamiltonian. One simply replaces each operator with the corresponding element of the color field



\begin{equation}
\begin{alignedat}{2}
H_{\textrm{SU(4)}} &= \lim_{M\to\infty}\left\langle \Psi \left\vert \mathcal{H} \right\vert \Psi \right\rangle\\
    &= \frac{J}{2} \sum_{j} \left(n^7_j + n^8_j + n^9_j \right) \\
    &+ J^{\prime} \sum_{j} \left(n^{1}_j n^{1}_{j+1} + n^{2}_j n^{2}_{j+1} + n^{3}_j n^{3}_{j+1}  \right) \\
    &+ J^{\prime} \sum_{j} \left(n^{4}_j n^{4}_{j+1} + n^{5}_j n^{5}_{j+1} + n^{6}_j n^{6}_{j+1}  \right).
\end{alignedat}
\label{eq:classical_hamiltonian}
\end{equation}
The classical dynamics is again derived by examining Heisenberg's equations of motion
\begin{equation}
\mathrm{i}\hbar\frac{d\hat{T}_{j}^{\alpha}}{dt} =\left[\hat{T}_{j}^{\alpha},\hat{\cal{H}}\right]
=\mathrm{i} f_{\alpha\beta\gamma}\frac{\partial\mathcal{\hat{H}}}{\partial\hat{T}_{j}^{\beta}}\hat{T}_{j}^{\gamma}, 
\label{eq:heisenberg_su4}
\end{equation}
where the operator derivative is used as a shorthand (see the derivation of the \gls*{LL} equations above) and $f_{\alpha\beta\gamma}$ are the anti-symmetric structure constants defined by the relation 
$\left[\hat{T}^\alpha, \hat{T}^\beta\right] = \mathrm{i}f_{\alpha\beta\gamma}\hat{T}^\gamma$. 
In terms of our basis Eq.~\eqref{eq:generators}, these are
\begin{equation}
\begin{alignedat}{2}
f_{1,2,3} = f_{1,8,11} = f_{2,9,13} = f_{2,11,14} = f_{3,7,15} &= 1, \\
f_{1,9,10} = f_{1,12,15} = f_{2,7,12} = f_{3,8,14} = f_{3,10,13} &= -1,\\
f_{4,5,6} = f_{4,8,10} = f_{5,9,12} = f_{5,10,15} = f_{6,7,14} &= 1, \\
f_{4,9,11} = f_{4,13,14} = f_{5,7,13} = f_{6,8,15} = f_{6,11,12} &= -1. \\
\label{eq:structure_constants}
\end{alignedat}
\end{equation}
Finally, we evaluate the expectation value of Eq.~\eqref{eq:heisenberg_su4} to arrive at the classical equations of motion
\begin{equation}
\frac{dn_{j}^{\alpha}}{dt}= \frac{f_{\alpha\beta\gamma}}{ \hbar}\frac{\partial H_{\rm SU(4)}}{\partial n_{j}^{\beta}} n_{j}^{\gamma}.
\label{eq:dimer_GLL}
\end{equation}

Equations~\eqref{eq:classical_hamiltonian},~\eqref{eq:structure_constants} and~\eqref{eq:dimer_GLL} together represent a generalization of the \gls*{LL} equations for SU(4) coherent states, following the program of Ref.~\cite{Zhang21}. We will refer to them as the \gls*{GLL} equations. We emphasize that these equations are completely general and do not depend on the choice of generators, which in turn determine the structure constants. This dynamics may in fact be cast in a more manifestly basis-independent fashion as a nonlinear Schr{\"o}dinger equation, a formulation that yields a number of conceptual and numerical advantages, as described in Appendix~\ref{sec:appendix_dynamics} and~\cite{Dahlbom22}.
The expanded set of observables used in this theory enables the representation of non-magnetic states as well
as singlet-triplet excitations. Moreover, these equations may be simulated numerically 
at a cost that is linear in the number of dimers. To demonstrate these features, and to determine the applicability
of the approximation as a whole, we will use the formalism of SU(4) coherent states to generate \gls*{DSSF} 
intensities $\mathcal{S}\left(\mathbf{q},\omega\right)$  and compare to zero temperature ($T=0$) \gls*{DMRG} simulations and finite-$T$ \gls*{ED} results. \\

\section{Estimation of $\mathcal{S}\left(\mb{q},\omega, {T}\right)$ \label{sec:S(q,w)}}
The classical \gls*{DSSF} can be calculated from dipole trajectories $s_j^\alpha(t)$ 
by Fourier transforming the associated spin-spin correlation function both in time and on the lattice
\begin{equation}
\mathcal{S}^{\alpha\beta}_{\rm{cl}}\left(\mb{q}, \omega, {T}\right) = \int_{-\infty}^{\infty} d\omega\hspace{1pt} e^{-\mathrm{i}\omega t} \left< s_{\mb{-q}}^\alpha(t) s_{\mb{q}}^\beta({0}) \right>,
\label{eq:dssf}
\end{equation}
where {$\langle ...\rangle$ indicates thermal average,} $\alpha$ and $\beta$ take values of $x$, $y$ or $z$ and
\begin{equation}
s_\mb{q}^\alpha = \frac{1}{\sqrt{N_s}}\sum_{j,\sigma} e^{\mathrm{i}\mb{q}\cdot\mb{r}_{j\sigma}} s_{j\sigma}^\alpha.
\label{eq:spin_lattice_fourier_transform}
\end{equation}
Here $j$ indexes rungs, $\sigma=l,r$ indicates the site on each rung, and $N_s$ is the total number of sites.

To calculate the DSSF of Eq.~\eqref{eq:dssf} with the dynamics presented in the previous section, one begins
by sampling an initial condition, $n_j^\alpha\left(0\right)$ from thermal equilibrium. This initial condition is
then evolved into a trajectory, $n_j^\alpha\left(t\right)$, using the dynamics of Eq.~\eqref{eq:dimer_GLL}. The spin-spin correlations may then be constructed by noting the correspondence between the first six observables given in Eq.~\eqref{eq:generators} and the familiar spin observables on each site.
While we clearly have the correspondence between 
$s_{j,l}^\alpha = n_j^\alpha$ and $s_{j,r}^\alpha = n_j^{\alpha+3}$, where $\alpha=1,2,3$, the
fact that $n_j^\alpha$ corresponds to the $l$ row of the lattice for $\alpha=1,2,3$ and to the $r$ row for $\alpha=4, 5, 6$ requires special treatment in calculation of the lattice Fourier transform. Consider an arbitrary wave vector, $\mb{q}=(q_x, q_y)$. If $L$ is the number of bonds and we assume
that the lattice constant along both axes is unity, then $q_x$ may take values of $2\pi n/L$, with $n=0,\ldots,L-1$,
and $q_y$ may take values of $0$ and $\pi$. Thus,
\begin{equation}
s_\mb{q}^\alpha = \frac{1}{\sqrt{2L}}\sum_{j=0}^{L-1} \sum_{\nu=0}^1 e^{\mathrm{i}q_xj}e^{\mathrm{i}q_y\nu} s_{j,\nu}^\alpha.
\end{equation}
We shall be particularly interested in the $q_y=\pi$ channel, i.e., the antisymmetric channel that carries information about single-triplon excitations on bonds. For this we have
\begin{equation}
s_{\mathbf{q}}^\alpha = \frac{1}{\sqrt{L}}\sum_{j=0}^{L-1} e^{\mathrm{i}q_xj} \frac{1}{\sqrt{2}}(s_{j,r}^\alpha - s_{j,l}^\alpha),
\end{equation}
In terms
of the variables of our generalized dynamics, the $\pi$ or antisymmetric channel of the dipolar dynamical structure factor can therefore be recovered from a generalized trajectory $n_j^\alpha(t)$ by 
calculating the correlations of $\left(n_j^{\alpha+3} - n_j^\alpha\right)/\sqrt{2}$ and performing the spatial Fourier transform only along the length of the ladder.

In the $T\to0$ limit, the dynamics of the classical equations of motion will produce trajectories
that are small oscillations about the classical ground states, 
which may be entirely characterized by a normal mode analysis. 
The quantization of the resulting normal modes gives a path to recovering the traditional \gls*{LSWT}, more typically arrived at by directly bosonizing the quantum Hamiltonian. It follows that the
``quantum'' \gls*{DSSF} predicted by \gls*{LSWT}, $\mathcal{S}_{\rm{Q}}^{\alpha\beta}\left(\mathbf{q}, \omega\right)$,
may be recovered from the classical calculation by observing that the respective correlation functions of the classical and quantum harmonic oscillators 
are related by a proportionality factor of
\begin{equation}
\frac{\hbar\omega}{k_B T}\left[1 + n_B\left(\hbar\omega/T\right)\right],
\label{eq:c2q_factor}
\end{equation}
where
$n_B\left(\hbar\omega/T\right)\equiv\left(e^{\hbar\omega/k_B T} - 1\right)^{-1}$ is the Bose function.
This leads to the conclusion that, in the $T\to 0$ limit,
\begin{equation}
\mathcal{S}_{\rm{Q}}^{\alpha\beta}\left(\bm{q}, \omega, {T}\right) = \frac{\hbar\omega}{k_B T} \left[1 + n_{\rm{B}}\left(\hbar\omega/T\right)\right] \mathcal{S}_{\rm{cl}}^{\alpha\beta}\left(\bm{q}, \omega, {T}\right),
\label{eq:Sq_from_Sc}
\end{equation}
as described in greater detail in \cite{dahlbom2024finiteT}. 

In the following section, we rely on the correspondence between the quantization of the normal modes of the classical theory and \gls*{LSWT} to study the applicability of our approach in the low-temperature regime. Specifically, we examine a generalization of \gls*{LSWT} that corresponds to our generalized classical dynamics. We will return to the purely classical dynamics in Sec.~\ref{sec:finite_T}.


\section{Zero Temperature Comparisons with Density Matrix Normalization Group \label{sec:T=0_DMRG_comparisons}}

\begin{figure*}
    \centering
    \includegraphics[width=1.0\textwidth]{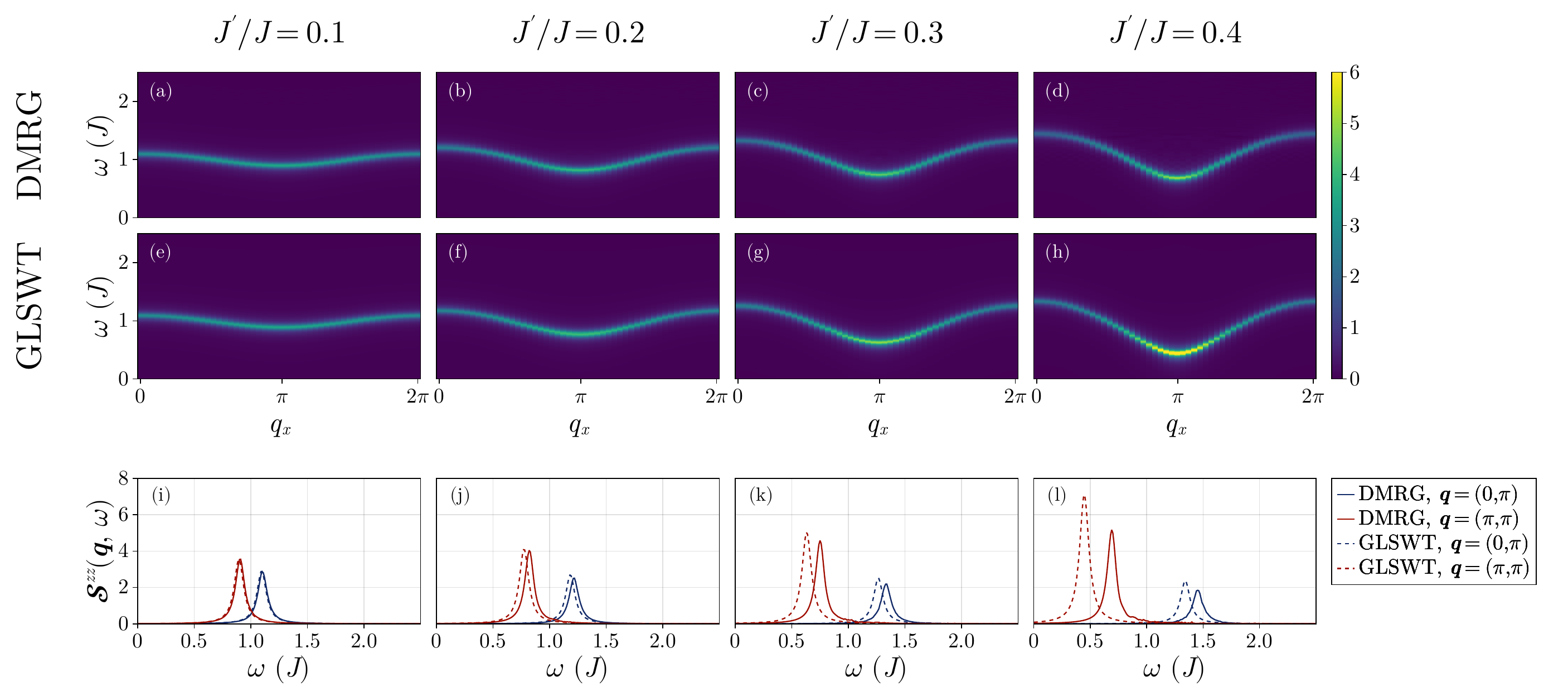}
    \caption{
    Structure factor intensities, $\mathcal{S}^{zz}\left(\mathbf{q},\omega\right)$, calculated using both the density-matrix renormalization group method [DMRG, panels (a)--(d)] and generalized linear spin wave theory [GLSWT, panels (e)--(h)]. The two top rows show the intensities along the reciprocal space path $\mathbf{q}=\left(q_x,\pi\right)$, with $q_x$ ranging from $0$ to $2\pi$. The bottom row [panels (i)--(l)] shows single-$\mathbf{q}$ cuts generated with each method at $\mathbf{q}=\left(0, \pi\right)$ and $\left(\pi, \pi\right)$. Columns correspond to different values of $J^\prime/J$.  
    } 
    \label{fig:DMRG_LSWT}
\end{figure*}

At low temperatures, the classical approach can be expected to break down for large enough $J^\prime/J$, in which case a proper description the ground state and excitations must take into account long range entanglement. To be more precise about this range of parameters, here we determine the $J^\prime/J$ values over which our methodology can approximately reproduce the results of \gls*{DMRG} in the $T\to0$ limit. As noted above, the classical dynamics of Eq.~\eqref{eq:classical_hamiltonian} reproduce those of \gls*{LSWT} in this limit after rescaling the intensities following Eq.~\eqref{eq:c2q_factor}. We therefore begin by calculating the quadratic spin wave Hamiltonian corresponding to the SU(4) classical theory outlined above. {This amounts to application of a multi-flavor bond-operator formalism, as is well-established in the literature \cite{matveev1973quantum, Sachdev1990, Chubukov1991, Normand2011, Toth2012}. We will refer to it as a \gls*{GLSWT} calculation. The novelty here consists solely in making a connection to the corresponding classical theory and identifying the appropriate control parameter for semiclassical expansions, $M$, which labels irreps of SU($4$). }

We have already identified a set of SU(4) generators, Eq.~\eqref{eq:generators}, for the purposes of deriving a classical theory. These also provide a natural language for rewriting our quantum Hamiltonian, Eq.~\eqref{eq:hamiltonian}, 
\begin{equation}
\hat{\mathcal{H}} = \frac{J}{2}\sum_{j;\alpha=7,9} \hat{T}_j^{\alpha} + J^{\prime}\sum_{j;\beta=1,6}\hat{T}_j^\beta\hat{T}_{j+1}^\beta 
\label{eq:hamiltonian_T_ops}
\end{equation}
We will recast this Hamiltonian in a bosonic language by introducing bosons, $b_{j,n}$, where $n=0,1,2,3$ labels flavor~\cite{muniz2014generalized}. We also introduce 
an explicit matrix representations of the generators, $\mathbb{T}^\beta$, in the standard, four-dimensional product basis: $\left\vert \uparrow\uparrow\right\rangle$, $\left\vert \uparrow\downarrow\right\rangle$, $\left\vert \downarrow\uparrow\right\rangle$, $\left\vert \downarrow\downarrow\right\rangle$. With ${\bm b}^{\dagger}_j \equiv (b^{\dagger}_{j,0}, b^{\dagger}_{j,1}, b^{\dagger}_{j,2}, b^{\dagger}_{j,3} )$, we then define a bosonic representation of the generators as
\begin{equation}
\mathcal{T}_j^\beta = {\mathbf b}^{\dagger}_j \mathbb{T}^{\beta} {\mathbf b}^{\;}_j.
\end{equation}
One may verify that the $\mathcal{T}^\beta_j$ satisfy identical commutation relations as our original generators
\begin{equation}
\left[\mathcal{T}_j^\alpha,\mathcal{T}_j^\beta\right] = \mathrm{i}f_{\alpha\beta\gamma}\mathcal{T}_j^\gamma,
\end{equation}
with $f_{\alpha\beta\gamma}$ defined as in Eq.~\eqref{eq:structure_constants}. The specific representation given by this construction is fixed by imposing the constraint
\begin{equation}
\sum_{n=0}^{3} b_{j,n}^\dagger b_{j,n} = M.
\end{equation}
where $M$ labels the completely symmetric irreps of SU(4) [just as $w$ labels irreps of SU(2)]. Setting $M=1$ corresponds to choosing the four-dimensional fundamental representation of interest. Note that this is precisely the same $M$ that is sent to infinity to derive the classical limit. 

We may now make the substitution $\hat{T}^\beta_j \to \mathcal{T}^\beta_j=\mathbf{b}_j^\dagger\mathbb{T}^\beta\mathbf{b}_j$ in Eq.~\eqref{eq:hamiltonian_T_ops} to arrive at an entirely equivalent Hamiltonian expressed in terms of the bosons $b_{j,m}$
\begin{equation}
\hat{\mathcal{H}} = \frac{J}{2}\sum_{j;\alpha=7,9} \mathbf{b}_j^\dagger\mathbb{T}^{\alpha}\mathbf{b}_j + J^{\prime}\sum_{j;\beta=1,6}\mathbf{b}_j^\dagger\mathbb{T}^\beta\mathbf{b}_j\mathbf{b}_{j+1}^\dagger\mathbb{T}^\beta\mathbf{b}_{j+1}.
\label{eq:hamiltonian_bosons_unrotated}
\end{equation}
We wish to approximate this Hamiltonian by expanding it in powers of $M$ and retaining only terms that are bilinear in the bosons. Prior to performing this expansion, however, we will set the local SU(4) quantization axis on each site along the direction of the local coherent state, $\left\vert\Psi_j\right\rangle$, of the classical ground state $\left\vert\Psi\right\rangle = \otimes_j\left\vert\Psi_j\right\rangle$. The classical ground state can be found by minimizing Eq.~\eqref{eq:classical_hamiltonian}. 
In the parameter region of interest ($J^\prime/J < 0.5$), the result, expressed in terms of the color field, is $n_j^\alpha=-\frac{1}{2}$ for $\alpha=7,8,9$ and $n_j^\alpha = 0$ for all other $\alpha$. Expressed as a complex vector in the product basis, we have the familiar singlet state $\left\vert \Psi_j \right\rangle = \left(0, 1/\sqrt{2}, -1/\sqrt{2},0\right)^T$, as can be determined by finding the highest-weight eigenstate of $n_j^\alpha \hat{T}_j^\alpha$ (with summation over $\alpha$). 

Since the matrices $\mathbb{T}^\beta$ were defined in the product basis, it follows that the $m=0$ bosons creates the fully polarized state from the vacuum state: $b_{j,0}\left\vert\varnothing\right\rangle = \left\vert\uparrow\uparrow\right\rangle$. We wish to find a transformed set of bosons containing a distinguished element, $b_{j,s}^\dagger$, that will instead create the singlet (ground) state from the vacuum state:
\begin{equation}
    \tilde{b}^{\dagger}_{j,s}\left\vert\varnothing\right\rangle = \left\vert \Psi_j\right\rangle.
\end{equation}
This can be achieved with a transformation, $U^\dagger$, the first column of which is the singlet state. The remaining columns may be chosen freely. For convenience we select the standard triplet basis elements for the $S=1/2$ dimer: $\left\vert\uparrow\uparrow\right\rangle$, $\left(\left\vert\uparrow\downarrow\right\rangle + \left\vert\downarrow\uparrow\right\rangle\right)/\sqrt{2}$, and $\left\vert\downarrow\downarrow\right\rangle$. With this transformation, we define a set of rotated bosons
\begin{equation} 
\tilde{\mathbf{b}}_{j} =  U_j \mathbf{b}_{j}, \quad \tilde{\mathbf{b}}_{j}^\dagger = \mathbf{b}_{j}^\dagger U_j^\dagger,
\label{eq:boson_transformation}
\end{equation}
where $\tilde{\bm b}^{\dagger}_j \equiv (\tilde{b}^{\dagger}_{j,s}, \tilde{b}^{\dagger}_{j,-1}, \tilde{b}^{\dagger}_{j,0}, \tilde{b}^{\dagger}_{j,1} )$. 
Note that we have changed the boson flavors from $\{0,1,2,3\}$ to $\{s,-1,0,1\}$ since $U_j$ was chosen such that $\tilde{b}^{\dagger}_{j,s}$
creates the singlet state of the dimer $j$ and the other three bosons, $\tilde{b}^{\dagger}_{j,m}$, create the triplet states with projections $m=-1,0,1$. Using the same transformation to define $\tilde{\mathbb{T}}_j^\alpha=U^{\phantom\dagger}_j\mathbb{T}^\alpha_jU_j^\dagger$, Eq.~\eqref{eq:hamiltonian_bosons_unrotated} may be rewritten equivalently as
\begin{equation}
\hat{\mathcal{H}} = 
 \frac{J}{2}\sum_{j;\alpha=7,9} \tilde{\mathbf{b}}_j^\dagger \tilde{\mathbb{T}}_j^{\alpha} \tilde{\mathbf{b}}_j
+ J^{\prime} \sum_{\substack{j; \beta=1,6  }}
\tilde{\mathbf{b}}_j^\dagger \tilde{\mathbb{T}}_j^\beta \tilde{\mathbf{b}}^{\phantom\dagger}_j \tilde{\mathbf{b}}_j^\dagger\tilde{\mathbb{T}}_{j+1}^\beta\tilde{\mathbf{b}}^{\phantom\dagger}_j. \\
\end{equation}

As a starting point for obtaining the spin wave Hamiltonian, we introduce the SU(4) generalization~\cite{muniz2014generalized} of the Holstein-Primakoff transformation
\begin{equation}
\tilde{b}_{j,s}^\dagger = \tilde{b}_{j,s} = \sqrt{M}\sqrt{1 - \frac{1}{M}\sum_{m=-1}^{1}\tilde{b}_{j,m}^\dagger \tilde{b}_{j,m}}.
\end{equation}
This transformation still provides a faithful representation of the SU(4) generators, but it gives us a route to an approximation via an expansion of the square root in powers of $M$.
Performing this expansion and collecting terms of order $M$ yields,
\begin{equation}
\begin{alignedat}{2}
\hat{\mathcal{H}}^{\left(2\right)} &= J\sum_{j; m=-1,1} \tilde{b}_{j,m}^\dagger\tilde{b}_{j,m} \\  &+J^\prime \sum_{j; m=-1,1} \left[\tilde{b}_{j,m}^\dagger\tilde{b}_{j+1,m} + {\rm H. c.} \right] \\
&+ J^\prime \sum_{j; m=-1,1}\left[ (-1)^m \tilde{b}_{j,m}\tilde{b}_{j+1, \bar{m}} + {\rm H. c.} \right],\\
\end{alignedat}
\end{equation}
where $\bar{m} \equiv - m$.
$\hat{\mathcal{H}}^{(2)}$ may be brought into block diagonal form by introducing a Fourier transform on the bosons. Since our local SU(4) coherent states encompass entire dimers, the irreps of the translation group are labeled by a single number, $q$, which corresponds to $q_x$ in our previous discussion. The Fourier transform is thus
\begin{equation}
\tilde{b}_{q,m}^\dagger = \frac{1}{\sqrt{L}} \sum_{j}e^{\mathrm{i}qj}\tilde{b}_{j,m}^\dagger,
\end{equation}
which yields
\begin{equation}
\begin{aligned}
\hat{\mathcal{H}}^{\left(2\right)}&= \sum_{q,m=-1,1}
\bigg[A_q \tilde{b}_{q,m}^\dagger\tilde{b}_{q,m} \\
&\quad\quad+
\frac{(-1)^m B_q}{2} \left(\tilde{b}_{q,m}\tilde{b}_{\bar{q},\bar{m}} 
+ \tilde{b}^\dagger_{\bar{q},\bar{m}}\tilde{b}^\dagger_{q,m} \right) \bigg]
\end{aligned}
\end{equation}
where $\bar{q} \equiv - q$ and
\begin{equation}
A_q =J + J^\prime\cos\left(q\right),\quad 
B_q =  J^{\prime}\cos\left(q\right).
\end{equation}
After eliminating anomalous terms using a Bogoliubov transformation, we arrive at the diagonal form of $\hat{\mathcal{H}}^{\left(2\right)}$ which, up to an irrelevant constant, is given by
\begin{equation}
\hat{\mathcal{H}}^{\left(2\right)} =   \sum_{q} \sum_{m=-1}^1 \omega\left(q\right) \beta^{\dagger}_{q,m} \beta^{\;}_{q,m},
\end{equation}
where the operators 
\begin{eqnarray}
\beta^{\dagger}_{q,\bar{1}} &=& \cosh{\theta_{q}} b^{\dagger}_{q,\bar{1}} - \sinh{\theta_{q}} b^{\;}_{q,1},
\nonumber \\
\beta^{\dagger}_{q,0} &=& \cosh{\theta_{q}} b^{\dagger}_{q,0} + \sinh{\theta_{q}} b^{\;}_{q,0},
\nonumber \\
\beta^{\dagger}_{q,1} &=& \cosh{\theta_{q}} b^{\dagger}_{q,1} - \sinh{\theta_{q}} b^{\;}_{q,\bar{1}},
\end{eqnarray}
with 
\begin{equation}
\tanh{2 \theta_q} = \frac{B_q}{A_q},
\end{equation}
create a collective triplon modes with momentum $q$ and eigenvalue $m$ of 
$\hat{S}^z_{\rm tot}= \sum_{j=1}^L (\hat{S}_{j,l}+ \hat{S}_{j,r}) $.
The corresponding triplon dispersion relation
\begin{equation}
\omega\left(q\right) =  J\sqrt{1 + \frac{2J^\prime}{J}\cos\left(q\right)},
\label{eq:dispersion_relation}
\end{equation}
does not depend on $m$ because the paramagnetic ground state does not break the SU(2) symmetry of the spin Hamiltonian. We note that the spin gap $\Delta$ is set by the minimum energy excitation:
\begin{equation}
\Delta = \omega\left(\pi\right) = J \sqrt{1 - \frac{2J^\prime}{J}}.
\label{eq:gap}
\end{equation}

In contrast with the known behavior of the ladder system, it is clear from Eq.~\eqref{eq:gap} that the dispersion becomes gapless when $J^\prime/J=0.5$ {(see Fig.~\ref{fig:J5_T0})}. To better determine the region of applicability of the semiclassical approach, we compare the predictions of both \gls*{GLSWT} and \gls*{DMRG} for values of $J^\prime/J = 0.1, 0.2, 0.3$ and $0.4$ on a system with $L=50$ bonds. The results are presented in Fig.~\ref{fig:DMRG_LSWT}. Details on the numerical calculations may be found in Appendix~\ref{sec:DMRG}.

Unlike traditional \gls*{LSWT}, which depends on an expansion in terms of $S$ [irreps of SU(2)] ($M=2S$), the \gls*{LSWT} based on an expansion in terms of $M$ [integer label of symmetric irreps of SU(4)] clearly captures the gapped triplon modes. Moreover, the qualitative features of the dispersion are apparent for all $J^\prime/J<0.5$. Significant quantitative departures began to manifest at $J^\prime/J=0.3$. In particular, the gap narrows too quickly in the semiclassical approach with a corresponding increase in intensity at bottom of the band. When intradimer exchange is substantially larger than the interdimer exchange, however, the efficacy of the approach is quite striking, particularly since this is a quasi-1D antiferromagnet being considered in the $T\to0$ limit. This suggests that suitably generalized classical and semiclassical approaches can be an effective and computationally tractable approach to modeling quantum magnets when the entanglement between sites is restricted to individual bonds or localized clusters of bonds.

\section{Finite Temperature Comparisons with Exact Diagonalization \label{sec:finite_T}}

The \gls*{GLSWT} approach described in Sec.~\ref{sec:T=0_DMRG_comparisons} is technique for treating magnets that exhibit only short-range entanglement in the $T\to 0$ limit. The full classical theory corresponding to \gls*{GLSWT}, presented in Sec.~\ref{sec:classical_theory}, opens the possibility of computationally inexpensive finite-$T$ calculations.
To facilitate this, Sec.~\ref{sec:finite-T_corrections} introduces a number of finite-temperature corrections that must be applied to the GLL calculations. In Sec.~\ref{sec:J'=0.1, 0.3 finite-T results}, we benchmark the results of the corrected GLL calculations to those of finite-temperature ED. This is done for $J^\prime/J=0.1$ and $0.3$, parameter values for which we have shown the generalized theory works well at $T=0$.

{Materials typically exhibit a crossover into a classical regime as the temperature exceeds some characteristic energy scale, $T^*$. One may then ask whether a classical approach can reproduce the correct high-temperature behavior even when it fails at $T=0$. In Sec.~\ref{sec:J'>=0.5} we address this question by studying the ladder model with $J^\prime/J=0.5$ and $0.7$, values which cause both the traditional and generalized classical theories fail at $T=0$. To test whether either of these classical theories -- LL or GLL -- exhibits the appropriate behavior in the high-temperature regime, we again benchmark against \gls*{ED}.}

\subsection{Finite-temperature corrections to $\mathcal{S}\left(\mathbf{q}, \omega, T\right)$ \label{sec:finite-T_corrections}}

Sec.~\ref{sec:S(q,w)} presented a general approach to estimating $\mathcal{S}_{\mathrm{cl}}^{\alpha\beta}\left(\bf{q}, \omega, T\right)$ in the $T\to0$ limit, relying
on the classical-to-quantum correspondence factor Eq.~\eqref{eq:c2q_factor}. At elevated temperatures, we introduce an additional correction
by enforcing the quantum sum rule through a systematic rescaling of expectation values. To motivate this idea, explained in more detail in \cite{dahlbom2024finiteT}, we begin by reviewing various sum rules that apply to quantum and classical structure factors. 

At $T=0$, the quantum mechanical \gls*{DSSF} is given by
\begin{equation}
\mathcal{S}^{\alpha\beta}_{\rm{Q}}\left(\mathbf{q}, \omega\right) = \sum_{\nu,\mu}\left\langle \nu \right\vert \hat{S}^\alpha_{-\mathbf{q}}\left\vert\mu\right\rangle \left\langle\mu\right\vert\hat{S}_{\mathbf{q}}^\beta\left\vert\nu\right\rangle \delta\left(\epsilon_\mu - \epsilon_\nu - \omega\right).
\label{eq:dssf_quantum}
\end{equation}
The trace of this expression satisfies a sum rule when integrated over all $\omega$ and all $\mathbf{q}$ in a single Brillouin zone: 
\begin{equation}
\int_{-\infty}^{\infty} \int_{\rm{BZ}}d\omega d\mathbf{q} \hspace{1pt} \sum_{\alpha=1}^3 \mathcal{S}_{\rm{Q}}^{\alpha\alpha}\left(\mathbf{q}, \omega\right) = N_{\rm{S}}S\left(S+1\right),
\label{eq:sum_rule_quantum}
\end{equation}
This follows from Parseval's relation and 
the quadratic Casimir of the Lie algebra $\mathfrak{su}\left(2\right)$, $\sum_\alpha \hat{S}^\alpha\hat{S}^\alpha=S\left(S+1\right)$, which we call $C^{(2)}_{\mathrm{SU}(2)}$.

The corresponding classical \gls*{DSSF} as defined in Eq.~\eqref{eq:dssf}, $\mathcal{S}_{\mathrm{cl}}^{\alpha\beta}\left(\mathbf{q}, \omega, T\right)$, also satisfies a sum rule:
\begin{equation}
\int_{-\infty}^{\infty} \int_{\rm{BZ}}d\omega d\mathbf{q} \hspace{1pt} \sum_{\alpha=1}^3\mathcal{S}_{\rm{cl}}^{\alpha\alpha}\left(\mathbf{q}, \omega, T\right) = N_{\rm{S}}S^2,
\label{eq:sum_rule_classical}
\end{equation}
which holds at all temperatures $T$. This follows from the fixed magnitude of classical dipoles in the SU(2) formalism, where the relevant
invariant is not the sum of square of the spin operators but the sum of the squares of the expectation values of the spin operators in any SU($2$) coherent state: $\sum_\alpha \left\langle\Omega\right\vert\hat{S}^\alpha\left\vert\Omega\right\rangle^2 = S^2$.
Note that the quantum sum rule Eq.~\eqref{eq:sum_rule_quantum} leads to the classical sum rule via the factorization rule upon taking the 
classical ($S \to \infty$) limit and keeping the leading order contribution $S^2$ in powers of $S$.

Because of the correspondence given in Eq.~\eqref{eq:Sq_from_Sc}, applying the classical-to-quantum correction factor of Eq.~\eqref{eq:c2q_factor} to $\mathcal{S}^{\alpha\beta}_{\textrm{cl}}$ prior to integration will result in approximate satisfaction of the quantum sum rule when $T$ is close to zero. As $T$ approaches infinity, however, the correspondence factor approaches 1 for all $\omega$, resulting in a reversion to the classical sum. 
Additionally, the justification for applying this factor breaks down at elevated temperatures, where the spin dynamics is not well described by a linear approximation (i.e., by harmonic oscillators). Motivated by these observations, we adopt another approach to enforcing the quantum sum rule in the high-temperature regime. 

In the $T\to\infty$ limit, the quantum sum may be recovered by rescaling all expectation values -- here those of the spin operators -- by the square root of the ratio of the quantum quadratic invariant, $S\left(S+1\right)$, to the classical one, $S^2$: $\sqrt{1+1/S}$. This renormalization was proposed in \cite{Huberman08} and has the effect of stiffening the system, 
maintaining an intensity distribution closer to quantum calculations at high temperatures. In between the low- and high-temperature limits, the quantum sum rule can be maintained by rescaling the expectation values by a factor, $\kappa$, that takes values between $1$ and $\sqrt{1+1/S}$. The exact value required is model dependent and may be determined empirically by simulation, as suggested in \cite{dahlbom2024finiteT}.

\begin{figure}[t]
    \centering
    \includegraphics[width=1\columnwidth]{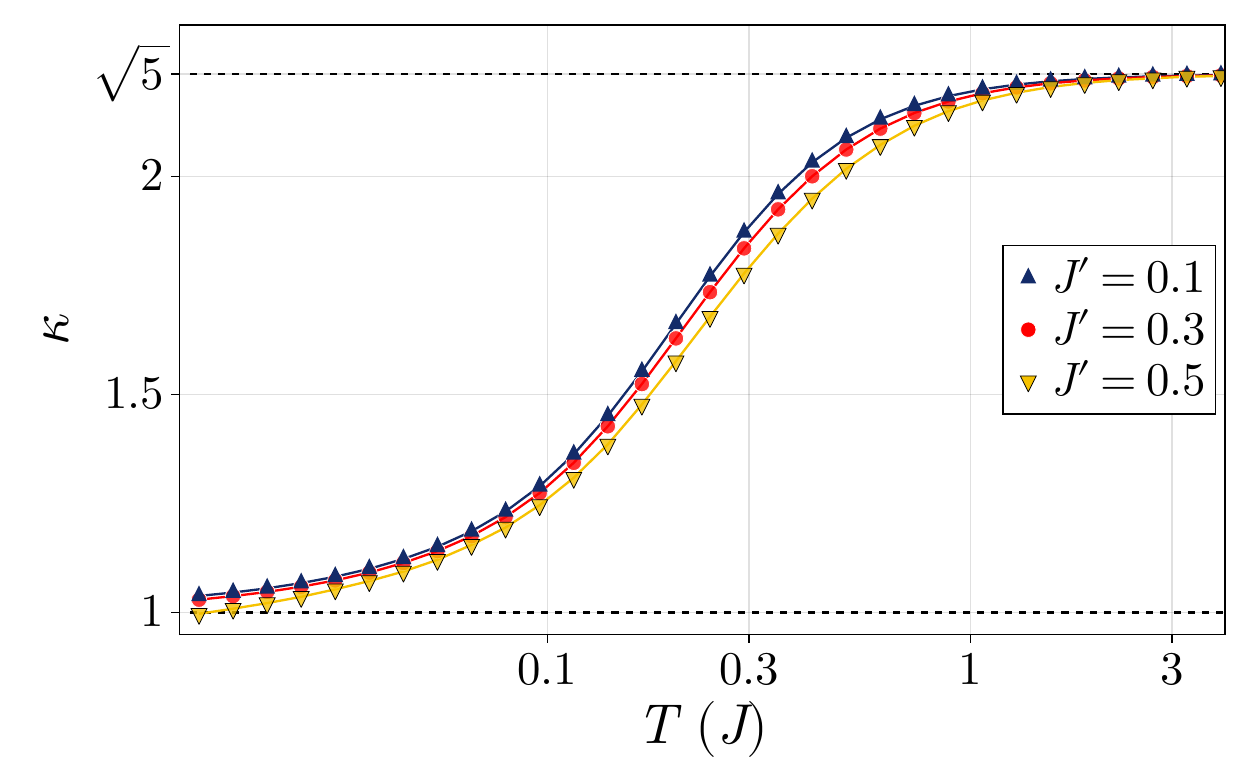}
    \caption{Renormalization of expectation values, $\kappa\left(T\right)$, applied to generalized spins $n^{\alpha}_j$ to ensure satisfaction of the quantum sum rule at each simulated temperature. At low temperatures, the 
    standard classical-to-quantum correction factor is sufficient to recover the sum rule, so $\kappa\left(0\right) = 1$. At high temperatures, the moment must be renormalized by the square root of the ratio of quantum to classical quadratic invariants of the Lie algebra $\mathfrak{su}\left( 4 \right)$, which in this case gives $\kappa\left(\infty\right)=\sqrt{5}$.}
    \label{fig:kappas}
\end{figure}

The approach just outlined may be extended to the classical dynamics of SU($4$) coherent states in a natural way. The generalization of the $T=0$ quantum \gls*{DSSF} is given by
\begin{equation}
\mathcal{T}^{\alpha\beta}_{\rm{Q}}\left(\mathbf{q}, \omega\right) = \sum_{\nu,\mu}\left\langle \nu \right\vert \hat{T}^\alpha_{-\mathbf{q}}\left\vert\mu\right\rangle \left\langle\mu\right\vert\hat{T}_{\mathbf{q}}^\beta\left\vert\nu\right\rangle \delta\left(\epsilon_\mu - \epsilon_\nu - \omega\right),
\end{equation}
where $\hat{T}^\alpha$ are generators of the group SU($4$), with $\alpha=1,\ldots,15$. We will employ the generators given in Eq.~\eqref{eq:generators}, which 
satisfy the quadratic Casimir $C^{(2)}_{\mathrm{SU}(4)}=\sum_\alpha \hat{T}^\alpha\hat{T}^\alpha = 15/4$. Consequently,
\begin{equation}
\int_{-\infty}^{\infty} \int_{\rm{BZ}}d\omega d\mathbf{q} \hspace{1pt} \sum_{\alpha=1}^{15} \mathcal{T}_{\rm{Q}}^{\alpha\alpha}\left(\mathbf{q}, \omega\right) = \frac{15}{4} L,
\label{eq:qsr}
\end{equation}
where $L$ is the number of rungs.

The classical \gls*{DSSF} is generalized by replacing the spin expectation values $s^\alpha = \left\langle \Omega \right \vert \hat{S}^\alpha \left\vert\Omega\right\rangle $ with elements of the color field introduced in Eq.~\eqref{eq:color_field}:
\begin{equation}
    \mathcal{T}_{\mathrm{cl}}^{\alpha\beta}\left(\mathbf{q},\omega, {T}\right) = \int_{-\infty}^{\infty} d\omega \hspace{1pt} e^{-\mathrm{i}\omega t}\left\langle n^{\alpha}_{-\mathbf{q}}\left(t\right) n^{\beta}_{\mathbf{q}}\left({0}\right)\right\rangle,
    \label{eq:dssf_generalized_classical}
\end{equation}
where $n^\alpha_{j}\left(t\right)$ is a classical trajectory generated by the dynamics of Eq.~\eqref{eq:dimer_GLL} and
\begin{equation}
n_\mathbf{q}^\alpha\left(t\right) = \frac{1}{\sqrt{L}} \sum_{j}e^{\mathrm{i}\mathbf{q}\cdot\mathbf{r}_j} n_j^\alpha\left(t\right).
\end{equation}
{{The brackets, $\left\langle\ldots\right\rangle$, again indicate thermal averaging.} The norm of the color field is $\sum_\alpha \left\langle\Psi\right\vert\hat{T}^\alpha\left\vert\Psi\right\rangle^2 = 3/4$, which holds for any SU($4$) coherent state $\left\vert\Psi\right\rangle$ and is 5 times smaller than the quantum sum rule~\eqref{eq:qsr}. Thus we have a classical sum rule corresponding to Eq.~\eqref{eq:sum_rule_classical}
\begin{equation}
\int_{-\infty}^{\infty} \int_{\rm{BZ}}d\omega d\mathbf{q} \hspace{1pt} \sum_{\alpha=1}^{15}\mathcal{T}_{\rm{cl}}^{\alpha\alpha}\left(\mathbf{q}, \omega, T\right) = \frac{3}{4} L.
\end{equation} 

\begin{figure*}
    \centering
    \includegraphics[width=1.0\textwidth]{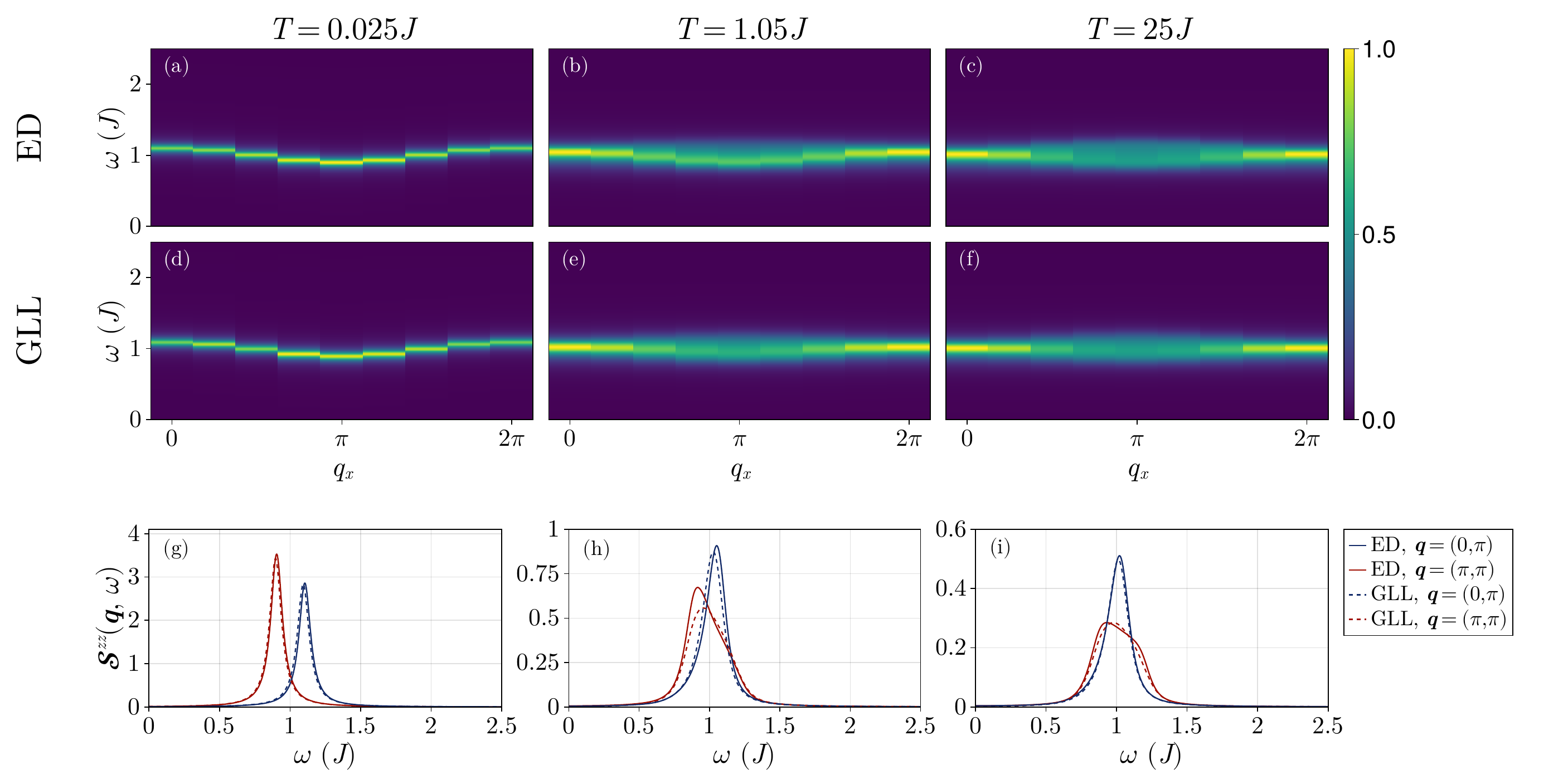}
   \caption{
    Structure factor intensities, $\mathcal{S}^{zz}\left(\mathbf{q},\omega, T\right)$, of the $J^\prime/J=0.1$ model calculated using both the exact diagonalization [ED, top rows, panels (a)--(c)] and the classical dynamics of Eq.~\eqref{eq:dimer_GLL} [\gls*{GLL}, middle row, panels (d)--(f)]. The contour plots show intensities along the reciprocal space path $\mathbf{q}=\left(q_x,\pi\right)$, with $q_x$ ranging from $0$ to $2\pi$. Each panel is individually normalized. The line plots [bottom row, panels (g)--(i)] show single-$\mathbf{q}$ cuts generated with each method at $\mathbf{q}=\left(0, \pi\right)$ and $\left(\pi,\pi\right)$. Columns correspond to different temperatures. 
    }
    \label{fig:ED_J1}
\end{figure*}

\begin{figure*}
    \centering
    \includegraphics[width=1.0\textwidth]{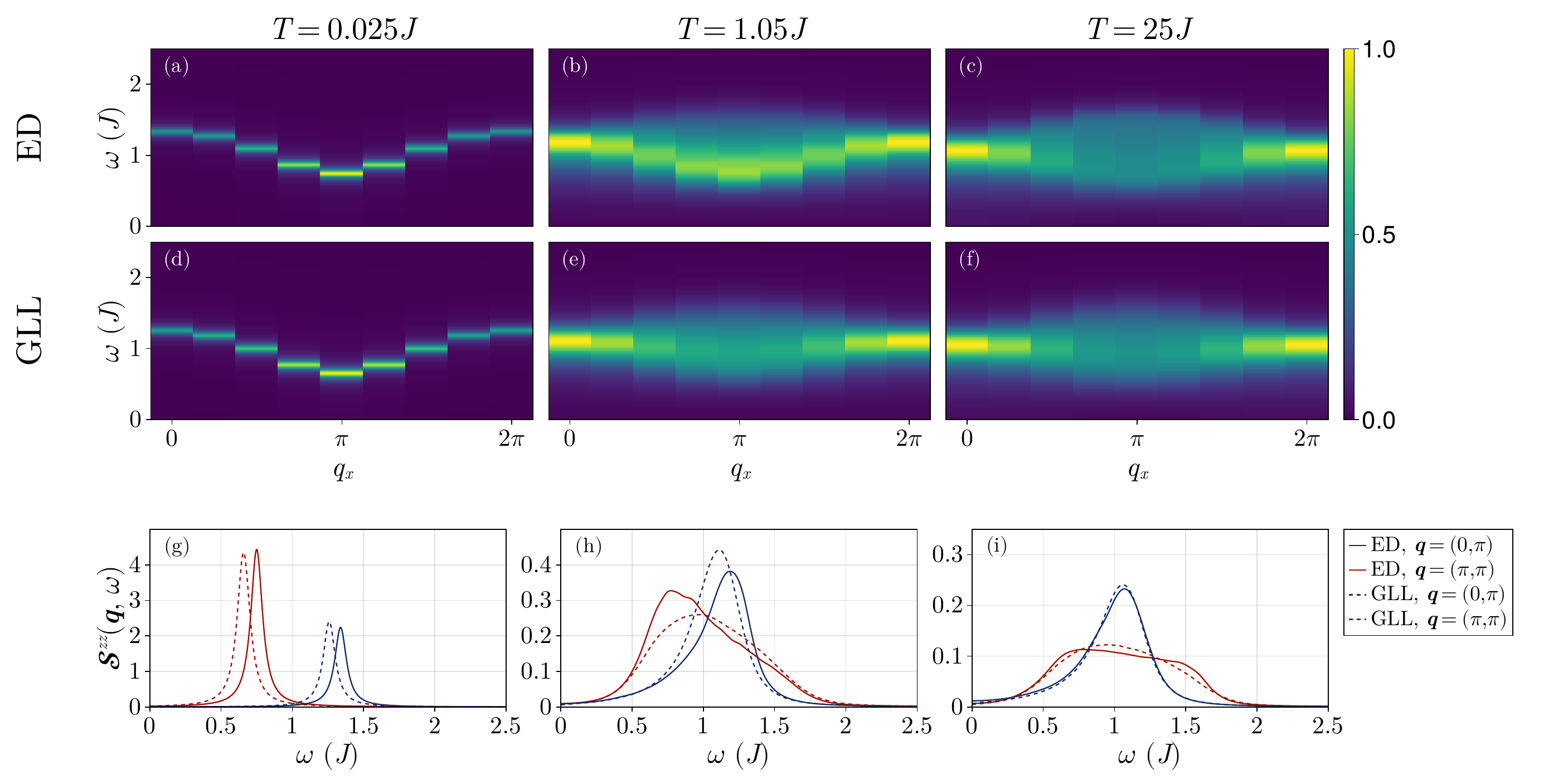}
    \caption{
    Structure factor intensities, $\mathcal{S}^{zz}\left(\mathbf{q},\omega, T\right)$, of the $J^\prime/J=0.3$ model calculated using both the exact diagonalization [ED, top rows, panels (a)--(c)] and the classical dynamics of Eq.~\eqref{eq:dimer_GLL} [\gls*{GLL}, middle row, panels (d)--(f)]. The contour plots show intensities along the reciprocal space path $\mathbf{q}=\left(q_x,\pi\right)$, with $q_x$ ranging from $0$ to $2\pi$. Each panel is individually normalized. The line plots [bottom row, panels (g)--(i)] show single-$\mathbf{q}$ cuts generated with each method at $\mathbf{q}=\left(0, \pi\right)$ and $\left(\pi,\pi\right)$. Columns correspond to different temperatures. 
    }
    \label{fig:ED_J3}
\end{figure*}
It follows that enforcing the quantum sum rule between the zero and infinite temperature limits demands application of a temperature-dependent moment renormalization factor, $\kappa\left(T\right)$, the values of which are bounded by $1$ and $\sqrt{5}$, respectively. For this study, $\kappa\left(T\right)$ was determined for $J^\prime=0.1,0.3$ and $0.5$ over a range of temperatures using an iterative procedure. The results are presented in Fig.~\ref{fig:kappas}. A full account of the calculation is given in \cite{dahlbom2024finiteT} and the implementation details are summarized in Appendix \ref{sec:appendix_kappa}. 

These renormalizations were applied to the dynamics of Eq.~\eqref{eq:dimer_GLL} by substituting 
\begin{equation}
n^\alpha_j\left(t\right) \rightarrow \kappa\left(T\right) n^\alpha_j\left(t\right).
\label{eq:moment_renormalization}
\end{equation}
The resulting trajectories were then used to estimate $\mathcal{S}_{\mathrm{cl}}^{\alpha\beta}\left(\mathbf{q},\omega, T\right)$ with the procedure described in Sec.~\ref{sec:S(q,w)}. As a reference, we also calculated the structure factor using \gls*{ED} and the Hamiltonian given in Eq.~\eqref{eq:hamiltonian}. To maintain a manageable computational cost for \gls*{ED}, we restricted the number of rungs to $L=8$ (corresponding to 16 spins). Finally, since the temperatures of the classical and quantum calculations cannot be expected to correspond, a linear rescaling was applied to the classical simulation temperatures. As there is no order parameter, and hence no ordering temperature to cross reference, the temperature rescaling was based on an analysis of the roll-off of intensity with increasing temperature. This rescaling factor is the only free parameter of our model that was empirically fit to ED data. The details are summarized in Appendix \ref{sec:appendix_temp_rescaling}. 

\subsection{Results for $J^\prime \leq 0.3J$} \label{sec:J'=0.1, 0.3 finite-T results}}

In Sec.~\ref{sec:T=0_DMRG_comparisons} we demonstrated that \gls*{GLSWT} provides an excellent approximation to \gls*{DMRG} results when $J^\prime\leq0.3$. We here examine whether the corresponding classical theory, the \gls*{GLL} dynamics described in Sec.~\ref{sec:classical_theory}, provides a good approximation at finite $T$. To this end, we estimated $\mathcal{S}_{\mathrm{cl}}\left(\mathbf{q}, \omega, T\right)$ at a range temperatures between $T = 0.015J$ and $25J$ and applied all 
the corrections described above: the classical-to-quantum correction factor Eq.~\eqref{eq:c2q_factor}, moment renormalization Eq.~\eqref{eq:moment_renormalization}, and temperature rescaling. These results were then compared to finite temperature \gls*{ED}.
(Complete details of the \gls*{ED} calculations are provided in Appendix~\ref{sec:appendix_ED}.)
Figs.~\ref{fig:ED_J1} and \ref{fig:ED_J3} show the respective results for $J^\prime/J=0.1$ and $0.3$ in three different temperature regimes: low ($T \ll J$), intermediate ($T \approx J$), and high ($T \gg J$).

We find that there is outstanding agreement between \gls*{ED} and the classical results with respect to the overall magnitude of the peak intensities and their distribution in momentum space. The broadening is also quantitatively similar for $q$ values close to 0. The chief discrepancies concern the shape of the peaks for $q$ values close to $\pi$. In particular, the \gls*{ED} results exhibit a more defined shoulder structure and a stronger skew toward lower energies, a phenomenon that becomes more pronounced for $J^\prime/J=0.3$. At very low temperatures, the $J^\prime/J=0.3$ results also exhibit a somewhat smaller gap than seen in the reference data, consistent with the \gls*{GLSWT} results reported in Fig.~\ref{fig:DMRG_LSWT}. We note, however, that the agreement becomes excellent again at high temperatures.

\subsection{ Results for $J^\prime \geq 0.5$ in the high-temperature regime \label{sec:J'>=0.5}}

\begin{figure}[t]
    \centering
    \includegraphics[width=1\columnwidth]{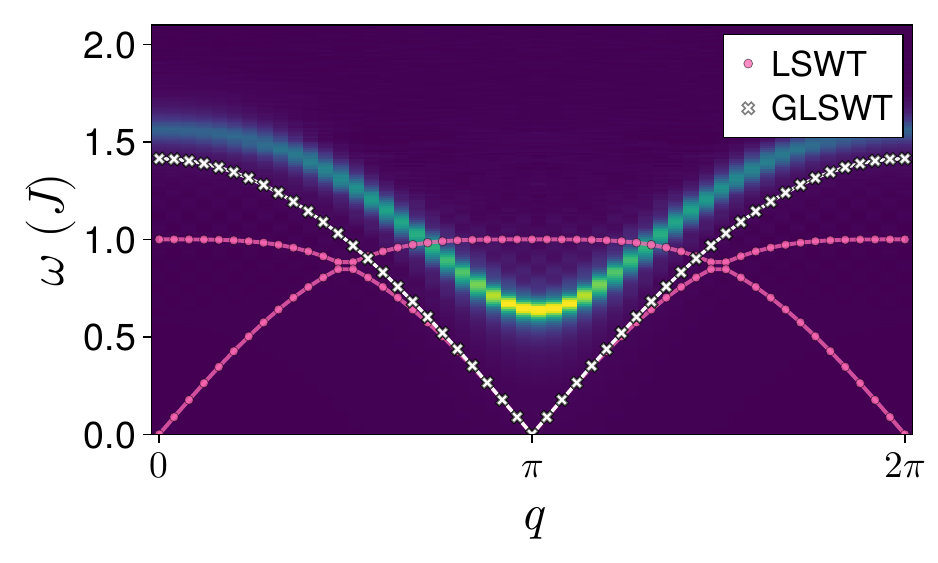}
    \caption{Zero temperature behavior of the $J^{\prime} = 0.5J$ spin ladder as modeled with three different methodologies. The background heatmap depicts the structure factor intensities as calculated with the density matrix renormalization group method (DMRG). The dispersions of both traditional linear spin wave theory (LSWT, pink circles) and generalized LSWT (GLSWT, white crosses) are overlaid. For values of $J^{\prime} = 0.5J$ and larger, both semiclassical approaches become gapless and fail to even qualitatively reproduce the correct dispersion. }
    \label{fig:J5_T0}
\end{figure}

As noted in Sec.~\ref{sec:T=0_DMRG_comparisons}, the classical ground state is no longer a pure singlet for values of $J^\prime /J \geq 0.5$. A finite dipole emerges on each site and antiferromagnetic order develops.  A \gls*{GLSWT} expansion about this ground state yields a Goldstone mode at $\mathbf{q}=[\pi,\pi]$, as illustrated in Fig.~\ref{fig:J5_T0}. Thus the semiclassical approach we have described fails both quantatively and qualitatively at $T=0$, as does traditional \gls*{LSWT}. Nonetheless, we can still ask whether the classical dynamics of Eq.~\eqref{eq:dimer_GLL} provides an effective description of the spin ladder in the high-temperature regime, where the length scale of entanglement effects is reduced and the system is expected to exhibit ``more classical'' behavior. 

\begin{figure*}
    \centering
    \includegraphics[width=1\columnwidth]{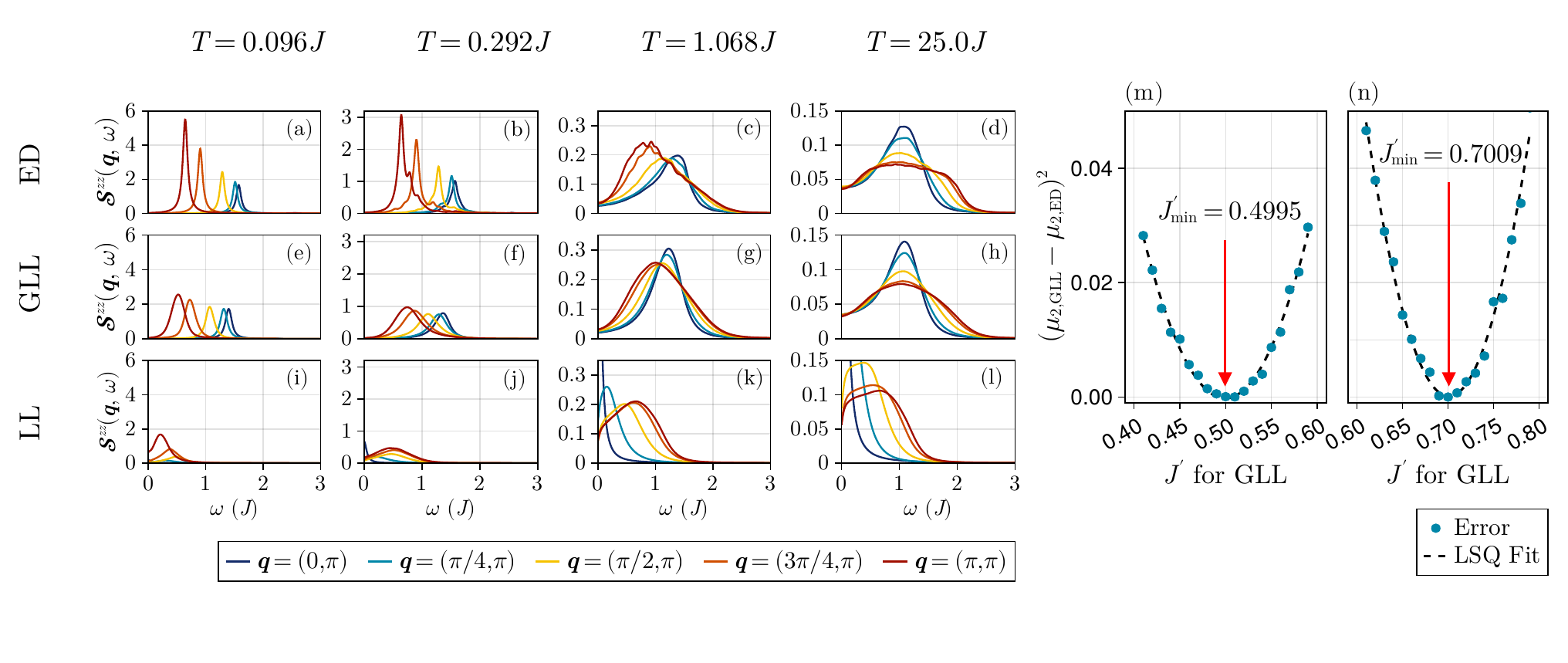}
    \caption{
    Panels (a)--(l) show the evolution of structure factor intensities, $\mathcal{S}^{zz}\left(\mathbf{q},\omega\right)$, with increasing temperature for $J^\prime =0.5J$, as modeled with exact diagonalization [ED, panels (a)--(d)], generalized Landau-Lifshitz dynamics [\gls*{GLL}, panels (e)--(h)], and traditional Landau-Lifshitz dynamics (\gls*{LL}, panels (i)--(l)). \gls*{ED} and \gls*{GLL} converge in the high temperature limit [panels (d) and (h)] while \gls*{LL} fails to capture the features of the \gls*{ED} calculation at all temperatures.
    Panels (m) \& (n) illustrate parameter extraction from the reference \gls*{ED} results using the \gls*{GLL} dynamics. The second moment of the \gls*{ED} result at $T=25 J$ was calculated for $J^\prime=0.5J$ (panel (m)) and $J^\prime=0.7J$ (panel (n)) as $\mu_{2,\mathrm{ED}}=\int_{0}^{3}\omega^2 \tilde{\mathcal{S}}^{zz}\left((\pi,\pi),\omega\right)d\omega$, where $\tilde{\mathcal{S}}^{zz}$ represents the normalized structure factor intensities. The second moments of the \gls*{GLL} results, $\mu_{2,\mathrm{GLL}}$, were calculated similarly for a range $J^\prime$ values. The figure shows the squared differences between the \gls*{ED} and \gls*{GLL} moments. By fitting a quadratic polynomial to these errors and finding the minimum, one obtains an excellent estimate of $J^\prime$ in the high temperature regime.
    }
    \label{fig:classical_crossover}
\end{figure*}

To explore this possibility, we calculated the structure factor intensities of the $J^\prime/J=0.5$ model at a range of temperatures using both \gls*{ED} and \gls*{GLL}, where the \gls*{GLL} simulations were again subjected to all the corrections described above. Additionally, we calculated the structure factor intensities using the traditional classical limit given by the \gls*{LL} equations. The \gls*{LL} simulations were also subjected to classical-to-quantum intensity rescaling and moment renormalization, but the temperature scale was left untouched -- the qualitative behavior of the \gls*{LL} results are so different from the other two that it is difficult to determine a clear correspondence. All results are given in Fig.~\ref{fig:classical_crossover}.

As expected, the low-temperature results of both the \gls*{GLL} and \gls*{LL} approaches do not agree well with the \gls*{ED} reference because the exact ground state is still a quantum paramagnet and the spectrum remains gapped. Notably, as shown in Fig.~\ref{fig:classical_crossover}(e), a pseudogap emerges in the \gls*{GLL} simulations, characterized by a depletion of spectral weight of $ \mathcal{S}^{zz}\left[{\bm q}=(\pi,\pi),\omega\right]$ below a temperature-dependent energy scale, whereas the \gls*{LL} approach maintains spectral weight extending down to zero energy [see  Fig.~\ref{fig:classical_crossover}(i)].
As $T$ approaches $J$, the results of both \gls*{ED} and \gls*{GLL} become qualitatively more similar, with both exhibiting an increasingly flat dispersion and similar broadening characteristics. This suggests that the crossover into the classical regime occurs at a temperature scale of $T^*\approx J$. In the high temperature regime, the agreement between \gls*{ED} and \gls*{GLL} becomes very good, similar to what was seen for $J^\prime/J=0.1$ and $0.3$. This is in contrast with the \gls*{LL} simulations, which are fundamentally incorrect throughout the entire temperature range.

The agreement between the quantum mechanical results and the generalized classical theory suggests an approach to extracting Hamiltonian parameters from inelastic scattering data collected at higher temperatures. As a simple illustration of this point, we calculate the second moment along the energy axis at the wave vector $\mathbf{q}=(\pi,\pi)$ of the \gls*{ED} data for $J^\prime/J=0.5$ and $0.7$
\begin{equation}
\mu_{2,\mathrm{ED}} = \int_{0}^{3} \omega^2 \tilde{\mathcal{S}}^{zz}((\pi,\pi),\omega)d\omega, 
\end{equation}
where $\tilde{\mathcal{S}}^{zz}$ is the normalized structure factor. We then
compare these to the second moment values estimated from the \gls*{GLL} results, $\mu_{2,\mathrm{GLL}}$, for a range of $J^\prime/J$ values. An excellent estimate of the true $J^\prime/J$ is achieved by fitting a quadratic polynomial to the squared differences between the \gls*{ED} moment and the \gls*{GLL} moments and then determining the minimum, as illustrated in Fig.~\ref{fig:classical_crossover} (m) and (n). We note that it is unnecessary to calculate $\kappa(T)$ to generate such an estimate, as the correct moment renormalization is known analytically in the high-temperature limit. Finally, we observe that this agreement persists beyond the critical point of $J^\prime/J=0.5$. The results for $J^\prime/J=0.7$, for example, are equally good.

Altogether, these results suggest that spin systems that exhibit high degrees of localized entanglement can still be studied classically in the high temperature regime. This demands application of the most appropriate classical theory, however, which we suggest is one based on entangled units, such as the dimer illustrated in this study. A significant benefit of the classical approach is the ability to extend to larger systems sizes at a costs that scales linearly with size, minimizing the finite-size effects that may substantially affect \gls*{ED} calculations and yielding a better estimate of system behavior in the thermodynamic limit.

\section{Discussion}
\label{sec:concl}

Deriving classical approaches that provide accurate approximations to the quantum dynamics of a spin system is crucial due to the significant difference in the computational costs between solving classical and quantum equations of motion. The numerical cost of solving classical equations scales linearly with the number of spins 
$N_\mathrm{s}$, making it computationally feasible even for large systems. In contrast, the cost of solving quantum equations of motion increases exponentially with $N_s$, making it prohibitively expensive for systems with many spins.

It is important to emphasize, however, that quantum spin systems admit multiple classical limits. While the standard $S \to \infty$ limit based on SU(2) coherent states is commonly used, it is fundamentally based on a decomposition of many-body state space into a product of 2-dimensional local Hilbert spaces. This restricts the possible classical ground states to dipoles. As a consequence, this approach is incapable of representing the non-magnetic ground states of quantum paramagnets, for example. Semiclassical \cite{muniz2014generalized} and classical techniques \cite{Zhang21, Dahlbom22, Dahlbom22b} based on SU($N$) coherent states incorporate any arbitrary local quantum state of an $N$-level unit
and have proven effective at overcoming this limitation for spin systems with $S>1/2$ by taking $N=2S+1$~\cite{Bai21, do2021decay, lee2023field, do2023understanding,  dahlbom2024finiteT}. Note that we use the word ``unit'' because  the content of each local Hilbert space need not be a single spin. Instead, each local space can represent a collection of entangled spins or both spin and orbit degrees of freedom. In this way the SU($N$) approach can also serve as a framework for analyzing systems composed of weakly-coupled $N$-level units.

In this study, we chose the quasi-1D system of a strong-rung 
$S=1/2$ spin ladder as a particularly stringent test case for this approach. We found that when the coupling within units is significantly greater than the coupling between units, classical and semiclassical methods provide a highly effective and computationally inexpensive approach to estimating a spin system's correlation functions.
The ground state closely resembles a direct product of singlets on each rung, deviating from the typical behavior under the traditional $S \to \infty$ limit. In such scenarios, resorting to coherent states from a different Lie group becomes necessary to derive classical (Lie-Poisson) equations of motion that can effectively model the low-temperature spin dynamics. A crucial question  that we addressed in this work is if such classical description also provides a good approximation away from the low-temperature regime.

While systems demonstrating long-range entanglement necessitate sophisticated approaches like \gls*{ED}, \gls*{DMRG}, or \gls*{QMC}, the methods outlined in this study are tailored to model many realistic materials where entanglement effects are mostly restricted to units compatible with lattice symmetries. Given that most materials exhibit a crossover to classical behavior at sufficiently high temperatures, our approach becomes particularly useful for estimating Hamiltonian parameters from high-temperature inelastic scattering data of materials showcasing exotic behavior at very low temperatures.

As demonstrated in the latter part of our study, Hamiltonian parameters can be estimated accurately by fitting the exact dynamical spin structure factor obtained via \gls*{GLL} dynamics in the infinite temperature limit. Once the model Hamiltonian is extracted using this classical approximation, it can be employed to investigate low-temperature dynamics using the more computationally intensive techniques mentioned earlier.

Accurate estimates of the dynamical spin susceptibility not only guide the search for new phases resulting from changes in Hamiltonian parameters but also serve as a tool to detect materials exhibiting strong inter-unit entanglement. This rational approach to the search for highly-entangled quantum states, such a quantum spin liquids, is proving to be effective for frustrated 2D magnets, such as triangular $J_1-J_2$ Heisenberg antiferromagnets~\cite{Ma16,Ito17,Ghioldi18,Macdougal20,Ghioldi22,Scheie23,Scheie24}. 

One clear limitation of classical approaches at low temperatures is their failure to reproduce the Bose-Einstein statistics of the collective modes. Recent works, as demonstrated in \cite{Dammak2009, Savin12, BarkerAndBauer2019}, have shown that this limitation can be overcome by substituting the typical white noise of Langevin dynamics with colored noise, which we leave for future investigation.

Finally, the generalized classical dynamics that we described here can be efficiently implemented via the Sunny package \cite{Sunny}, which greatly facilitates its application to real materials and the computation of measurable quantities, such as the temperature dependence of the inelastic neutron scattering cross-section.

\acknowledgments

This work was supported by the  U.S.~Department of Energy, Office of Science, Office of Basic Energy Sciences, under Award Number DE-SC0022311. This manuscript has been authored by UT-Batelle, LLC, under contract DEAC05-00OR22725 with the US Department of Energy (DOE). The US government retains and the publisher, by accepting the article for publication, acknowledges that the US government retains a nonexclusive, paid-up, irrevocable, worldwide license to publish or reproduce the published form of this manuscript, or allow others to do so, for US government purposes. DOE will provide public access to these results of federally sponsored research in accordance with the DOE Public Access Plan (http://energy.gov/downloads/doepublic-access-plan).

\begin{appendices}

\section{Density Matrix Renormalization Group Calculations}\label{sec:DMRG}
Our \gls*{DMRG} calculations for the \gls*{DSSF} are performed in real space using the DMRG++~\cite{alvarez0209} code and employing a center site approximation. Specifically, we compute the real-space correlation function 
\begin{multline}\label{eq:Srw}
    \mathcal{S}^{zz}_{(j,\sigma),(j^\prime,\sigma^\prime)}(\omega + i\eta) = \\
    \quad-\frac{1}{\pi}  \ \mathrm{Im} \bra{\Psi_\text{GS}} \hat{S}_{j,\sigma}^z \frac{1}{\omega - \mathcal{H} + E_\text{GS} + i \eta} \hat{S}_{j^\prime,\sigma^\prime}^z   \ket{\Psi_\text{GS}}, 
\end{multline}
where $j$ and $\sigma$ specify any site and $j^\prime=L/2-1$ and $\sigma^\prime=r$ together represent the reference site at the center of the cluster. $\ket{\text{GS}}$ is the ground state of the Hamiltonian with energy $E_\text{GS}$, and $\eta=0.05$ is an artificial broadening parameter. $\mathcal{S}^{zz}_{(j,\sigma),(j^\prime,\sigma^\prime)}(\omega)$ is then Fourier transformed into momentum space using periodic boundary conditions~\cite{DMRGKrylov}
\begin{align}
\mathcal{S}^{zz}_{\rm{Q}}\left(\mathbf{q}, \omega\right) &= \sum_{j=0}^{L-1} \cos{(q_x(j - L/2+1))} \notag \\ &\times \Bigg[ \mathcal{S}^{zz}_{(j,r),(j^\prime,\sigma^\prime)}(\omega + i\eta)  
- \mathcal{S}^{zz}_{(j,l),(j^\prime,\sigma^\prime)}(\omega + i\eta) \Bigg].
\end{align}

All \gls*{DMRG} calculations were performed on $50 \times 2$ clusters. For the ground state calculation, we keep between $m=200$ and $2000$ states to maintain a maximum truncation error of $10^{-10}$. For the \gls*{DSSF} calculations, we employ the root$-\mathcal{N}$ correction vector (CV) algorithm~\cite{rootN} based on the Krylov space method~\cite{DMRGKrylov} and compute Eq.~\eqref{eq:Srw} for each value of $\omega$. We set $\mathcal{N}=8$ and $\eta = J/20$ with an energy step of $\Delta \omega=J/10$, and keep between $m=200$ and $2000$ states to maintain a maximum truncation error of $10^{-8}$. 

\section{Exact Diagonalization Calculations}\label{sec:appendix_ED}
We performed finite temperature \gls*{ED} calculations on $8 \times 2$ clusters. Since the $z$-component of the total spin operator is a good quantum number for the many-body eigenstates, we fully diagonalize the Hamiltonian in magnetization sectors $\mathcal{{M}}_z= -7$ to $\mathcal{{M}}_z= 7$. The \gls*{DSSF} is then computed for each value of $T$ and $J$ using 
\begin{equation}
\mathcal{S}^{zz}_{\rm{Q}}\left(\mathbf{q}, \omega\right) = \frac{1}{\mathcal{Z}} \sum_{\nu,\mu} e^{-\epsilon_\nu\beta} 
|\mathcal{M}_{\mu,\nu}({\bf q})|^2\delta\left(\epsilon_\mu - \epsilon_\nu - \omega\right),
\end{equation}
where 
\begin{equation}
    |\mathcal{M}_{\mu,\nu}({\bf q})|^2=\left\langle \nu \right\vert \hat{S}^z_{\mathbf{q}}\left\vert\mu\right\rangle \left\langle\mu\right\vert\hat{S}_{-\mathbf{q}}^z\left\vert\nu\right\rangle. 
\end{equation}
Here, $\mathcal{Z} = \sum_\nu e^{-\epsilon_\nu\beta}$ is the partition function, $\beta = 1/T$ is the inverse temperature, and $\left\vert\nu \right\rangle$ are the many-body eigenstates of the Hamiltonian with energy $\epsilon_\nu$. 

\section{Schr{\"o}dinger Formulation of Classical Dynamics\label{sec:appendix_dynamics}} 

All calculations in this and the following appendices were performed with the Sunny package \cite{Sunny}, which provides an implementation of the generalized dynamics derived of Sec~\ref{sec:classical_theory} in the Schr{\"o}dinger picture \cite{Dahlbom22}. In this formulation, one introduces complex $N$-vectors, $\mathbf{Z}_j$, to represent the SU($N$) coherent states on each site $j$. For the ladder problem, we take $N=4$, consider each rung as a ``site'', and use the standard product basis $\left\vert\uparrow\uparrow\right\rangle$, $\left\vert\uparrow\downarrow\right\rangle$, $\left\vert\downarrow\uparrow\right\rangle$, and $\left\vert\downarrow\downarrow\right\rangle$. It can then be shown that the dynamics of Eq.~\eqref{eq:dimer_GLL} is entirely equivalent to the canonical Hamiltonian dynamics
\begin{equation}
\frac{d}{dt}\mathbf{Z}_j = -\mathrm{i}\frac{\partial H_{\mathrm{SU}(N)}}{\partial \mathbf{Z}_j^\ast}.
\end{equation}
This can equivalently be written
\begin{equation}
\frac{d}{dt}\mathbf{Z}_j = -\mathrm{i}\,\frak{H}_j \mathbf{Z}_j, 
\end{equation}
where
\begin{equation}
\frak{H}_j = \frac{\partial H_{\mathrm{SU}(N)}}{\partial n_j^\alpha} \mathbb{T}^\alpha
\end{equation}
is a local, time-varying Hamiltonian and the $\mathbb{T}^\alpha$ are the generators of Eq.~\eqref{eq:generators} expressed in the product basis. The expectation values
\begin{equation}
n_j^\alpha = \mathbf{Z}_j^\dagger \mathbb{T}^\alpha \mathbf{Z}_j
\end{equation}
are now explicitly dependent on the coherent states. An advantage of this approach is that the SU($N$) Casimirs are exactly conserved, and resulting dynamics in $\mathbf{Z}_j$ have a canonical Hamiltonian structure. The latter makes possible numerical integration schemes that are exactly symplectic (no energy drift). See \cite{Dahlbom22} for details.

\section{Classical Structure Factor Calculations\label{sec:appendix_kappa}}

For each temperature $T$, we sampled $10,000$ equilibrium spin configurations, $n_{j,n}^\alpha$, where $j$ is the rung index, $\alpha$ the generalized spin component, and $n$ is the sample index. The samples were generated by numerically solving a Langevin equation for the generalized spin dynamics of Sec.~\ref{sec:classical_theory}. This was done with a second-order Heun integration scheme using the Schr{\"o}dinger formulation of the dynamics, as described in \cite{Dahlbom22b}. The Langevin dynamics requires setting an empirical damping parameter, $\lambda$, which governs coupling to the thermal bath. For the results reported here, $\lambda=0.4$ was chosen for the purposes of balancing a rapid decorrelation time with a reasonable step size. A time step of $\Delta t= 0.01\hspace{2pt}J^{-1}$ was determined to be numerically stable for this choice of coupling coefficient and the range of temperatures examined. The system was initialized into a product of pure singlets and then thermalized for $30 \hspace{2pt} J^{-1}$, a duration sufficient to achieve ergodicity of the energy time-series at all the tested temperatures. 

The resulting spin configurations were then rescaled by $\kappa\left(T\right)$ and used as an initial condition for a trajectory $n_{j,n}^\alpha\left(0\right) = \kappa\left(T\right)n_{j,n}^\alpha$. The full trajectories $n_{j,n}^\alpha\left(t\right)$ were then evolved according to the dynamics of Eq.~\eqref{eq:dimer_GLL}. These equations were solved using the symplectic (dissipationaless) numerical scheme described in \cite{Dahlbom22}. A time step of $\Delta t=0.08\hspace{2pt}J^{-1}$ was selected for numerical stability. Samples of the numerical trajectory were taken every 13 steps, resulting in an effective time step of $\Delta t=1.04\hspace{2pt}J^{-1}$. 601 such samples were collected. Each trajectory was then Fourier transformed on the lattice and in time using the FFTW package to yield $n_{q,n}^\alpha \left(\omega\right)$. Here $q$ takes values $n\pi/4$ with $n=-3,\ldots,4$ and $\omega$ takes values $n \pi / \Delta t$ with $n=-300,\ldots,300$, or 601 evenly-spaced bins lying approximately between $-3$ and $3\hspace{2pt}J$. The convolution theorem then enables estimation of the generalized structure factor of Eq.~\eqref{eq:dssf_generalized_classical} as
\begin{equation}
\mathcal{T}^{\alpha\beta} = \frac{1}{N_{\mathrm{samples}}} \sum_{n} n^\alpha_{\mathbf{q}, n}\left(\omega\right) n^\beta_{-\mathbf{q},n}\left(-\omega\right).
\end{equation}
After applying the harmonic classical-to-quantum correction factor Eq.~\eqref{eq:c2q_factor} to this result, the total spectral weight was estimated as
\begin{equation}
I_{\mathrm{total}} = \int_{\infty}^{\infty}\int_{\mathrm{BZ}} d\omega d\mathbf{q} \frac{\omega}{k_B T}\left[1 + n_{B}\left(\omega/T\right)\right]\sum_{\alpha=1}^{15} \mathcal{T}^{\alpha\alpha}\left(\mathbf{q},\omega\right), 
\end{equation}
where $\hbar = 1$. Since we were working with discrete-time trajectories, this last expression reduces to a simple sum.

By construction, $I_{\mathrm{total}}$ will lie between values of $\frac{3}{4}N_{s}$ (classical sum rule) and $\frac{15}{4}N_s$ (quantum sum rule), where $N_s$ is the number of sites (see Sec.~\ref{sec:finite_T}). The goal is to find a $\kappa\left(T\right)$ that ensures satisfaction of the quantum sum rule to an acceptable tolerance. To achieve this, an initial guess of $\kappa\left(T\right)$ was chosen and the procedure outlined above was performed. If the resulting $I_{\mathrm{total}}$ was too small (too large), a larger (smaller) kappa was selected and the process repeated. To accelerate this process, a standard binary search algorithm was implemented. The process was stopped when the sum rule was satisfied to within an absolute tolerance of $0.001$. The results are presented in Fig.~\ref{fig:kappas}.

Both the DMRG and ED calculations included an intrinsic broadening parameter $\eta$~\eqref{eq:Srw}. When making comparisons to these results, all classical structure factors, $\mathcal{S}_{\mathrm{cl}}\left(\mathbf{q},\omega\right)$, were convolved with a Lorentzian kernel along the energy axis
\begin{equation}
\mathcal{S}_{\mathrm{cl,broad}}\left(\mathbf{q},\omega\right) = \int_{-\infty}^{\infty}d\tilde{\omega} L\left(\tilde{\omega}\right)\mathcal{S}_{\mathrm{cl}}\left(\mathbf{q}, \omega-\tilde{\omega}\right),
\end{equation}
where
\begin{equation}
L\left(\omega\right) = \frac{\eta^2}{\omega^2 - \eta^2}.
\end{equation}

\section{Temperature rescaling \label{sec:appendix_temp_rescaling}}

\begin{figure*}ķ    \centering
    \includegraphics[width=1\textwidth]{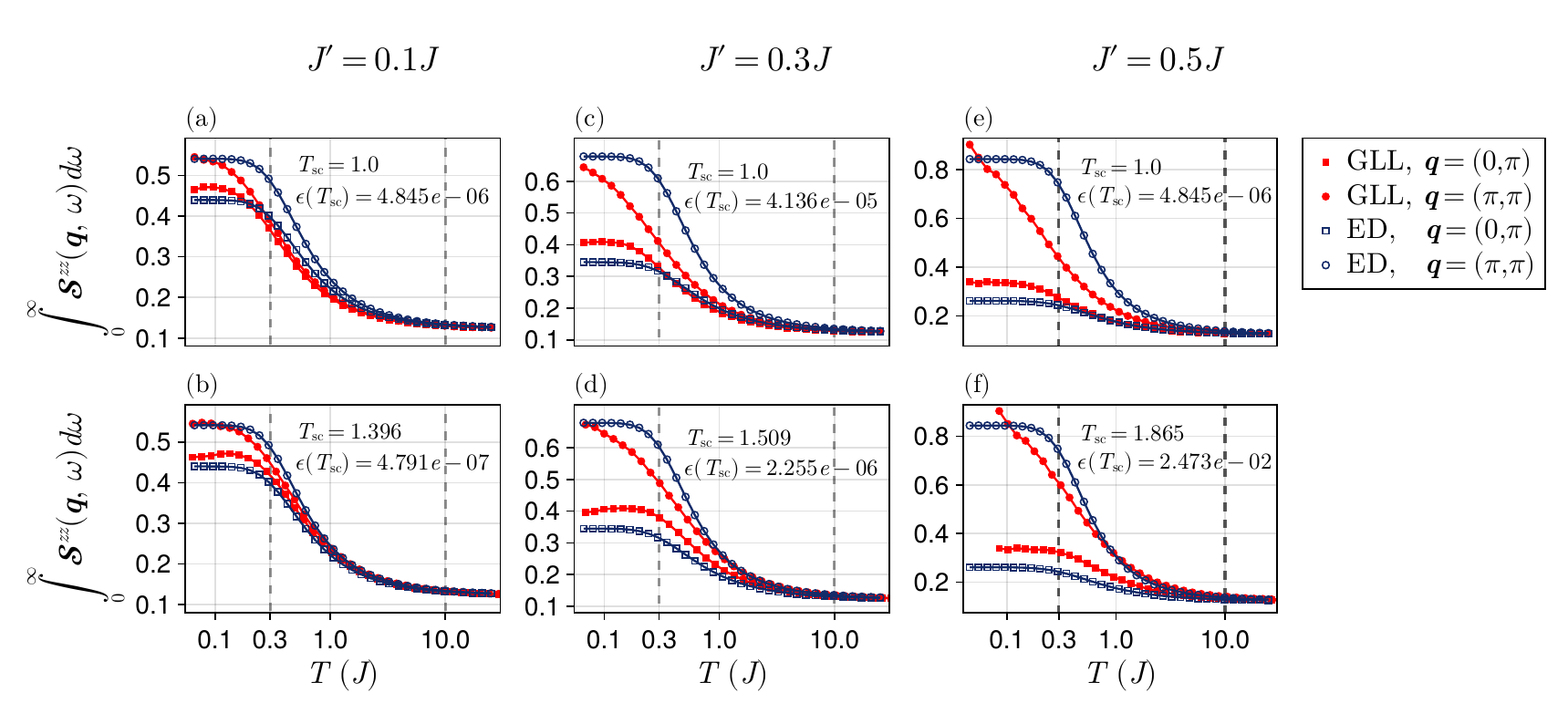}
    \caption{Integrated intensities, $\int_0^{\infty}\mathcal{S}^{zz}\left(q,\omega\right)d\omega$, at the wave vectors $\mathbf{q}=(0,\pi)$ and $\mathbf{q}=(\pi, \pi)$ as calculated with the generalized Landau-Lifsthiz dynamics (\gls*{GLL}, red) and exact diagonalization (\gls*{ED}, dark blue). For each value of $J^\prime$, the top figure shows the results without temperature rescaling and the bottom shows the results with temperature rescaling. The dashed lines indicate the region over which the error was evaluated. $T_{\mathrm{sc}}$ refers to the temperature rescaling factor, and $\epsilon\left(T_{\mathrm{sc}}\right)$ is the corresponding error.}
    \label{fig:temp_rescaling}
\end{figure*}

The temperature scale of a classical calculation is expected to deviate from the corresponding quantum calculation due
to the fundamentally different thermodynamic properties of each model. It is often possible to propose an \emph{ad hoc} rescaling of the classical temperature to put it into correspondence with the quantum system by examining the behavior of the order parameter in each case. For example, the classical temperature may be rescaled so that the Néel temperatures of both approaches correspond. We have no order parameter for the dimer ladder, however, as we are primarily concerned with values of $J^\prime/J$ for which the ground state is paramagnetic (product of singlets). 

The different low-temperature thermodynamic properties of the quantum and classical systems arise from the qualitative difference between
the Bose-Einstein and Boltzmann statistics. For example, examining the temperature dependence of the integrated intensity at a given wave vector, the quantum mechanical result for $I_Q \left(T,\mathbf{q} \right)-I_Q \left(0,\mathbf{q} \right)$ exhibits an activated behavior
($\propto e^{-\Delta/k_B T}$) due to the gapped nature of the spectrum. In contrast, the classical result for $I_{\rm cl} \left(T,\mathbf{q} \right)-I_{\rm cl} \left(0,\mathbf{q} \right)$ exhibits a power law behavior.
This qualitative discrepancy can potentially be remedied with a more sophisticated treatment of the thermal bath used in the classical simulations, as described in the references \cite{Dammak2009, Savin12, BarkerAndBauer2019, Franke2022}, but we do not pursue that approach here.

Given the characteristic differences between classical and quantum systems at low temperatures and given the lack of a distinguished ordering wave vector, we chose to determine the temperature rescaling by fitting the integrated intensity versus temperature curves at two wave vectors, $q=0$ and $q=\pi$, and do this in the transition and high-temperature regions only. Specifically, we calculated the integrated intensity at these two wave vectors and at 41 logarithmically-spaced temperatures between $T=0.015J$ and $25J$. We calculated these intensities with both the classical dynamics of Sec.~\ref{sec:classical_theory} and with \gls*{ED}. We fit the result of each approach with with cubic splines and refer to the resulting
interpolants as $I_{\mathrm{cl}}\left(T,\mathbf{q}\right)$ and $I_{\mathrm{ED}}\left(T,\mathbf{q}\right)$ respectively, where $\mathbf{q}$ is either $\mathbf{q}_1=(0,\pi)$ or $\mathbf{q}_2 = (\pi, \pi)$. The error was defined as
\begin{equation}
\begin{alignedat}{2}
\epsilon\left(T_{\mathrm{sc}}\right) = \int_{0.3}^{10.0}dT & \left(\left\vert I_{\mathrm{cl}}\left(T/T_{\mathrm{sc}}, \mathbf{q}_1\right) - I_{\mathrm{Q}}\left(T, \mathbf{q}_1\right)\right\vert  \right.\\
 &\left. - \left\vert I_{\mathrm{cl}}\left(T/T_{\mathrm{sc}}, \mathbf{q}_2\right) - I_{\mathrm{Q}}\left(T,\mathbf{q}_2\right)\right\vert\right)^2, 
\end{alignedat}
\end{equation}
where the bounds $T=0.3J$ and $T=10.0J$ were chosen to capture the transition region and portions of the high-temperature tail, and $T_{\mathrm{sc}}$ is the rescaling factor applied to the classical temperature. $\epsilon\left(T_{\mathrm{sc}}\right)$ was then minimized numerically. The resulting $T_{\mathrm{sc}}$ was taken as the rescaling temperature for all calculations in this paper. This process was performed for $J^\prime/J=0.1, 0.3$ and $0.5$. The results are summarized in Fig.~\ref{fig:temp_rescaling}.

\end{appendices}

\bibliography{Refs}
\end{document}